# Quantum computation and complexity theory

Course given at the Institut für Informationssysteme,
Abteilung für Datenbanken und Expertensysteme,
University of Technology Vienna,

Wintersemester 1994/95

K. Svozil

Institut für Theoretische Physik

University of Technology Vienna

Wiedner Hauptstraße 8-10/136

A-1040 Vienna, Austria

e-mail: svozil@tph.tuwien.ac.at

December 5, 1994

qct.tex

**Abstract**

*The Hilbert space formalism of quantum mechanics is reviewed with emphasis on applications to quantum computing. Standard interferomeric techniques are used to construct a physical device capable of universal quantum computation. Some consequences for recursion theory and complexity theory are discussed.*

# Contents

# 1 The Quantum of action

"Quantization," as it is presently understood, has been introduced by Max Planck[1] in 1900 [65] in an attempt to study the energy spectrum of blackbody radiation.[2] "Quantization," according to Planck, is the *discretization* of the total energy $U_N$ of $N$ linear oscillators ("Resonatoren"),

$$U_N = P\varepsilon \in \{0, \varepsilon, 2\varepsilon, 3\varepsilon, 4\varepsilon, \ldots\}, \tag{1}$$

where $P \in \mathbb{N}_0$ is zero or a positive integer. $\varepsilon$ is the *smallest quantum of energy*. It is a linear function of frequency $\nu$ and proportional to Planck's fundamental constant $h$; i.e.,

$$\varepsilon = h\nu. \tag{2}$$

In 1905, Einstein's *light quantum hypothesis* extended Planck's ("Resonator") quantization to the electromagnetic field [29]. In Einstein's own words (cf. [29], p. 133),[3]

> Es scheint nun in der Tat, daß die Beobachtungen über die "schwarze Strahlung", Photoluminiszenz, die Erzeugung von Kathodenstrahlen durch ultraviolettes Licht und anderer Erzeugungung bzw. Verwandlung des Lichtes betreffende Erscheinungsgruppen besser verständlich erscheinen unter der Annahme, daß die Energie des Lichtes diskontinuierlich im Raume verteilt sei. Nach der hier ins Auge zu fassenden Annahme ist bei der Ausbreitung eines von einem Punkte ausgehenden Lichtstrahles die Energie nicht kontinuierlich auf größer und größer werdende Räume verteilt, sondern es besteht dieselbe aus einer endlichen Zahl von in Raumpunkten lokalisierten Energiequanten, welche sich bewegen, ohne sich zu teilen und nur als Ganze absorbiert und erzeugt werden können.

Einstein's light quantum hypothesis asserts that, as far as emission and absorption processes are concerned, the energy of a light ray which is emitted at some point is not distributed continuously over increasing regions of space, but is concentrated in a finite number of energy quanta, which can only be absorbed and emitted as a whole. With this assumption, the photoelectric effect could properly be described (cf. Figure 1).

---

[1] After some earlier proposals which failed [64], Planck arrived at the assumption of a discretization of energy levels associated with a particular oscillator frequency by the careful analysis of all derivation steps leading to the experimentally obtained form of the blackbody radiation.

[2] The energy spectrum is the distribution of energy over the frequencies at a fixed temperature of a "black body." A "black body" is thereby defined as any physical object which is in internal equilibrium. Assume that absorprtion and reflection processes play a minor rôle. Then, depending on the temperature, but irrespective of its surface texture, a "black body" will appear to us truly black (room temperature), warm red (3000 Kelvin), sun-like (5500 K), blue (> 7000 K).

[3] It indeed seems to be the case that the observations of blackbody radiation, photoluminescence, the generation of cathode rays by ultraviolet light and other phenomena related to the generation and annihilation of light, would become better understandable with the assumption that the energy of light is distributed discontinuously in space. According to the assumption proposed here, the radiation energy of light from a point source is not spread out continuously over greater and greater spatial regions, but instead it consists of a finite number of energy quanta which are spatially localized, which move without division and which can only be absorbed and emitted as a whole.

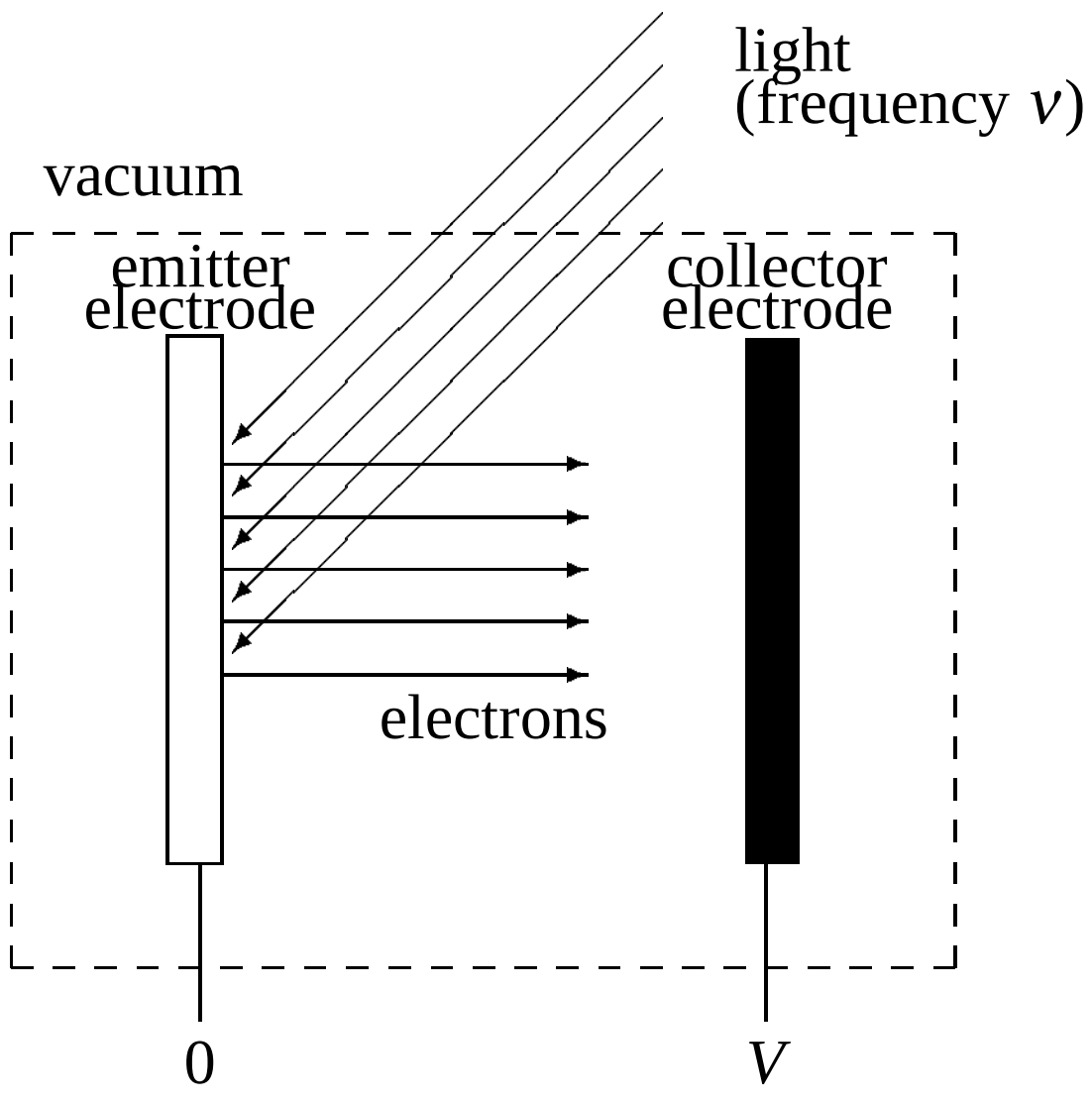

Figure 1: The photoelectric effect. A beam of light of frequency $\nu$ impinges upon an electrode. Electrons are emitted with kinetic energy $E = h\nu - W$, where $W < h\nu$ is the "threshold energy" necessary to release an electron (it is material dependent). Any increase in the intensity of light of frequency $\nu$ is accompanied by an increase in the number of emitted electrons of energy $h\nu - W$. The energy of the single electrons is *not* altered by this intensity increase. An increase in the Frequency $\nu' > \nu$, however, yields an energy increase of the emitted electrons by $\Delta\nu = \nu' - \nu$.

Thus, in extension of Planck's discretized resonator energy model, Einstein proposed a quantization of the electromagnetic field. Every field mode of frequency $\nu$ could carry a discrete number of light quanta of energy $h\nu$ per quantum (cf. 2.2, p. 15).

The present quantum theory is still a continuum theory in many respects: for infinite systems, there is a continuity of field modes of frequency $\omega$. Also the quantum theoretical coefficients characterising the mixture between orthogonal states, as well as space and time and other coordinates remain continuous — all but one: action. Thus, in the old days, discretization of phase space appeared to be a promising starting point for quantization. In a 1916 article on the structure of physical phase space, Planck emphasized that the quantum hypothesis should not be interpreted at the level of energy quanta but at the level of action quanta, according to the fact that the volume of $2f$-dimensional phase space ($f$ degrees of freedom) is a positive integer of $h^f$ (cf. [66], p. 387),[4]

> Es bestätigt sich auch hier wieder, daß die Quantenhypothese nicht auf Energieelemente, sondern auf Wirkungselemente zu gründen ist, entsprechend dem Umstand, daß das Volumen des Phasenraumes die Dimension von $h^f$ besitzt.

Since position and momentum cannot be measured simultaneously with arbitrary accuracy, the classical notion of a point in phase space has to be substituted by the notion of a cell of volume $h^f$. Stated differently: for periodic, onedimensional systems, the area of phase space occupied by the $n$'th orbit is

$$\int\int dp\, dq = \int p\, dq = nh. \tag{3}$$

Let us consider two examples: a linear oscillator and the quantum phase space of a rotator [77]. For the onedimensional linear oscillator with frequency $\nu$, the equation of motion is

$$q(2\pi\nu)^2 + \frac{d^2q}{dt^2} = 0. \tag{4}$$

Equation (4) has the solution $q = a \sin 2\pi\nu t$, where $a$ is an arbitrary constant. The canonical momentum is $p = m\frac{dq}{dt} = 2\pi\nu m a \cos 2\pi\nu t$. Elimination of the time parameter $t$ yields a trajectory in the $(p, q)$-phase space which is an ellipse; i.e., $\frac{q^2}{a^2} + \frac{p^2}{b^2} = 1$, where $b = 2\pi\nu m a$. The area of the ellipse is $ab\pi = 2\pi^2\nu m a^2$. Insertion of (3) yields

$$2\pi^2\nu m a^2 = nh. \tag{5}$$

Figure 2a shows the area of phase space occupied by the $n \leq 10$'th orbit for a onedimensional harmonic oscillator with $a = 2b$.

As a second example, consider a *rotator*, defined by a constant circular motion of a mass $m$ and of radius $a$ around a center. Let $q = \varphi$ be the angular coordinate in the plane of motion, then the rotator energy is given by

$$E = \frac{m(a\dot{q})^2}{2}. \tag{6}$$

[4] Again it is confirmed that the quantum hypothesis is not based on energy elements but on action elements, according to the fact that the volume of phase space has the dimension $h^f$.

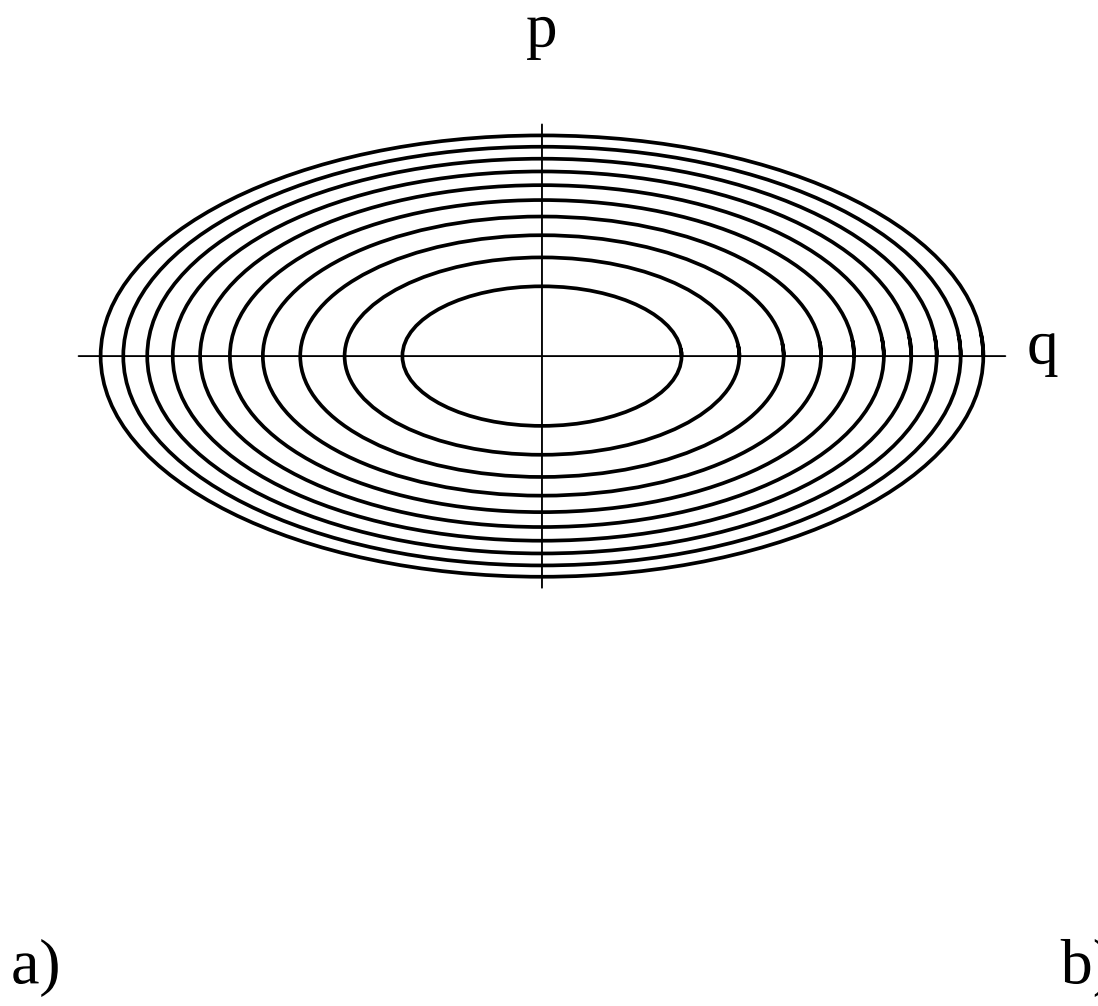

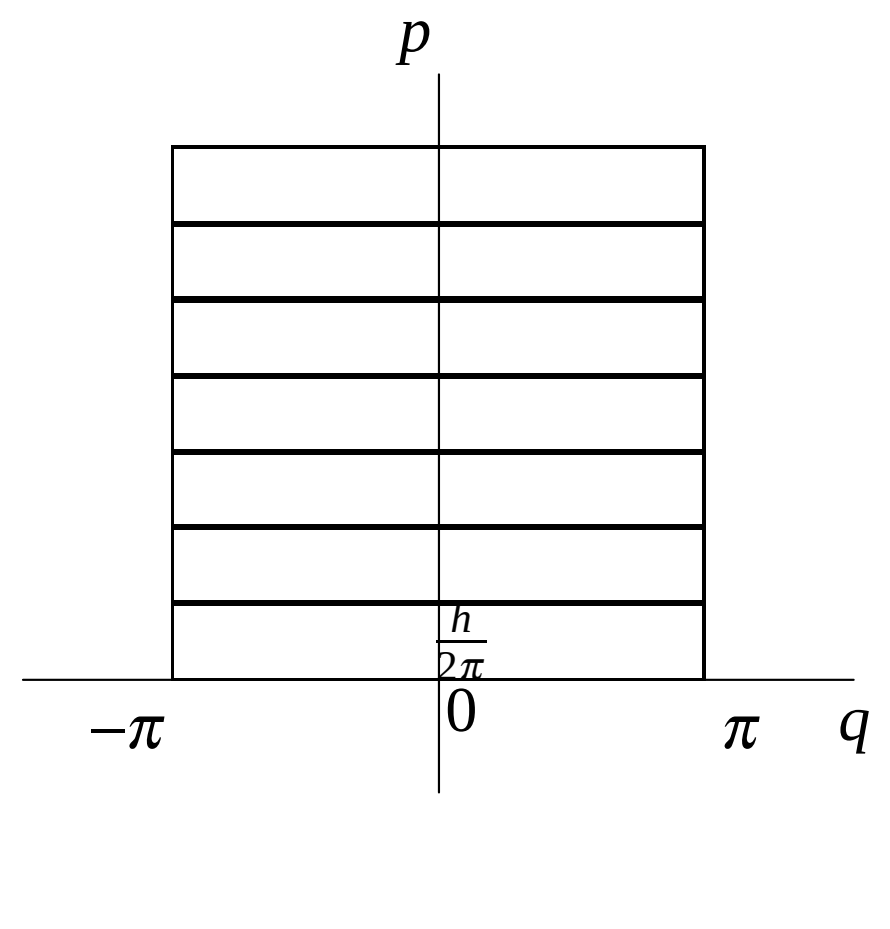

Figure 2: a) area of phase space occupied by the $n \leq 10$'th orbit for a onedimensional harmonic oscillator. b) area of phase space occupied by the first orbits of a rotator.

The associated momentum is $p = \frac{dE}{d\dot{q}} = ma^2\dot{q}$. For constant motion, $\dot{q}$ is constant and $-\pi < q \leq \pi$, one obtains by equation (3) a quantization of momentum of the form

$$2\pi p = nh. \tag{7}$$

Figure 2b shows the area of phase space occupied by the first orbits of a rotator.

Now, then, what does the quantum mean? Is it merely a metaphor, a way to compute? Or is it an indication of a discrete organization of the physical universe? One may safely state that the rôle of the quantum and our understanding of it as a hint towards a more fundamental discrete theory has not changed much over the years. As Einstein puts it ([31], p. 163),[5]

> Man kann gute Argumente dafür anführen, daß die Realität überhaupt nicht durch ein kontinuierliches Feld dargestellt werden könne. Aus den Quantenphänomenen scheint nämlich mit Sicherheit hervorzugehen, daß ein endliches System von endlicher Energie durch eine *endliche* Zahl von Zahlen (Quanten-Zahlen) *vollständig* beschrieben werden kann. Dies scheint zu einer Kontinuums-Theorie nicht zu passen und muß zu einem Versuch führen, die Realität durch eine rein algebraische Theorie zu beschreiben. Niemand sieht aber, wie die Basis einer solchen Theorie gewonnen werden könnte.

[5]There are good reasons to assume that nature cannot be represented by a continuous field. From quantum theory it could be inferred with certainty that a finite system with finite energy can be *completely* described by a *finite* number of (quantum) numbers. This seems not in accordance with continuum theory and has to stipulate trials to describe reality by purely algebraic means. Nobody has any idea of how one can find the basis of such a theory.

# 2 Quantum mechanics for the computer scientist

Is there a difference between quantum theory for physicists and quantum theory for logicians and computer scientists? Of course not, in principle!

Yet, a second glance reveals that there is a difference in aim. Courses in "hard-core" quantum theory for physicists tend to stress potential theory, and there the two solvable problems — the hydrogen atom and the harmonic oscillator. Computer scientists are more interested in quantum information and computing. They would like to concentrate on quantum coherence and the superposition principle and are therefore more attracted by recent developments in the "foundations" of quantum mechanics. (A very few are even attracted by quantum logic; for mere curiosity, it seems!) The following brief outline attempts to satisfy this demand.

## 2.1 Hilbert space quantum mechanics

In what follows, we shall make a great leap in time [43, 84], thereby omitting the Schrödinger-de Broglie wave mechanics and Heisenberg's formalism, and consider "state-of-the-art" Hilbert space quantum mechanics [28, 83, 58, 32, 40, 2, 50, 63]. (For a short review of Hilbert spaces, see appendix A.) It consists of the following (incomplete list of) rules. Thereby, every physical entity of a quantized system corresponds to an object in or defined by Hilbert space.

*(I)* Following Dirac [28], a physical *state* is represented by a *ket vector* of complex Hilbert space $\mathfrak{H}$, or *ket,* represented by the symbols "$|\rangle$". In order to distinguish the kets from each other, a particular letter or index or other symbol is inserted. Thus, the vector $\psi \in \mathfrak{H}$ is represented by the symbol "$|\psi\rangle$".

Since kets are defined as vectors in complex Hilbert space, any linear combination of ket vectors is also a ket vector. I.e.,

$$|\psi\rangle = a\,|1\rangle + b\,|2\rangle \in \mathfrak{H} \quad , \quad a,b \in \mathbb{C} \quad . \tag{8}$$

For continuous index $t$ and infinite dimensional Hilbert space,

$$|\psi\rangle = \int_{t_1}^{t_2} a(t)\,|t\rangle \in \mathfrak{H} \quad , \quad a(t) \in \mathbb{C} \quad . \tag{9}$$

Such a linear combination of states is also called "coherent superposition" or just "superposition" of states.

*(II)* Vectors of the dual Hilbert space $\mathfrak{H}^\dagger$ are called *bra vectors* or *bras.* They are denoted by the symbol "$\langle|$". Again, in order to distinguish the bras from each other, a particular letter or index or other symbol is inserted. Thus, the vector $\psi \in \mathfrak{H}^\dagger$ is represented by the symbol "$\langle\psi|$".

The metric of $\mathfrak{H}$ can now be defined as follows. Assume that there is a one-to-one correspondence (isomorphy) between the kets and the bras. Bra and ket thus corresponding to each other are said to be conjugates of each other and are labelled by the same symbols. Thus,

$$| \psi\rangle = (\langle\psi |)^\dagger \quad , \qquad \langle\psi | = (| \psi\rangle)^\dagger \quad . \tag{10}$$

Here, the symbol "†" has been introduced to indicate the transition to dual space, with the following syntactic rules:

$$(a)^\dagger \rightarrow a^* \tag{11}$$
$$(\langle\psi |)^\dagger \rightarrow | \psi\rangle \tag{12}$$
$$(| \psi\rangle)^\dagger \rightarrow \langle\psi | \tag{13}$$
$$(\langle\psi | \varphi\rangle)^\dagger \rightarrow \langle\varphi | \psi\rangle = (\langle\psi | \varphi\rangle)^* \quad . \tag{14}$$

Note that, by this definition, $(a \mid 1\rangle + b \mid 2\rangle)^\dagger = a^*\langle 1 \mid + b^*\langle 2 \mid$, where "∗" denotes complex conjugation.

*(III)* The *scalar product* of the ket $| \psi\rangle$ and the ket $| \varphi\rangle$ is the number $\langle\varphi \mid \psi\rangle$, i.e., the value $\varphi(| \psi\rangle)$ taken by the linear function associated with the bra conjugate to $| \varphi\rangle$.[6]

*(IV)* Elements of the set of orthonormal base vectors $\{| i\rangle \mid i \in \mathbb{I}\}$ ($\mathbb{I}$ stands for some index set of the cardinality of the dimension of the Hilbert space $\mathfrak{H}$) satisfy

$$\langle i | j\rangle = \delta_{ij} = \begin{cases} 0 & \text{if } i \neq j \\ 1 & \text{if } i = j \end{cases} \quad . \tag{15}$$

where $\delta_{ij}$ is the Kronecker delta function. For infinite dimensional Hilbert spaces, $\delta_{ij}$ is substituted by the Dirac delta function $\langle x \mid y\rangle = \delta(x-y) = \frac{1}{2\pi}\int_{-\infty}^{\infty} e^{i(x-y)t}dt$, which has been introduced for this occasion.

Furthermore, any state $| \psi\rangle$ can be written as a linear combination of the set orthonormal base vectors $\{| i\rangle \mid i \in \mathbb{I}\}$; i.e.,

$$| \psi\rangle = \sum_{i\in\mathbb{I}} a_i \mid i\rangle \tag{16}$$

with

$$a_i = \langle i \mid \psi\rangle \in \mathbb{C} \quad . \tag{17}$$

The identity operator **1** (not to be confused with the index set!) can be written in terms of the orthonormal basis vectors as ($a_i = 1$)

$$\mathbf{1} = \sum_{i\in\mathbb{I}} | i\rangle\langle i | \quad . \tag{18}$$

The sums become integrals for continuous spectra.

E.g., if the index $i$ is identified by the spatial position (operator) $x$ and the state is $| \psi(t)\rangle$ time-dependent, then $\langle x \mid \psi(t)\rangle = \psi(x,t)$ is just the usual (Schrödinger) wave function.

[6] Thereby, "||=|".

*(V) Observables* are represented by self-adjoint operators $\hat{R} = \hat{R}^\dagger$ on the Hilbert space $\mathfrak{H}$. For finite dimensional Hilbert spaces, bounded self-adjoint operators are equivalent to bounded Hermitean operators. They can be represented by matrices, and the self-adjoint conjugation is just transposition and complex conjugation of the matrix elements.

Self-adjoint operators have a spectral representation

$$\hat{R} = \sum_n r_n \hat{P}_n \quad , \tag{19}$$

where the $\hat{P}_n$ are orthogonal projection operators related to the orthonormal eigenvectors of $\hat{R}$ by

$$\hat{P}_n = \sum_a \mid a, r_n\rangle\langle a, r_n \mid \quad . \tag{20}$$

Here, the $r_n$ are the eigenvalues of $\hat{R}$, and the parameter $a$ labels the degenerate eigenvectors which belong to the same eigenvalue of $\hat{R}$. For nondegenerate eigenstates, equation (19) reduces to $\hat{R} = \sum_n r_n \mid r_n\rangle\langle r_n \mid$. Again, the sums become integrals for continuous spectra. Note also that, because of self-adjointness, self-adjoint operators in complex Hilbert space have real-valued eigenvalues; i.e., $r_n \in \mathbb{R}$.

For example, in the base $\{\mid x\rangle \mid x \in \mathbb{R}\}$, the position operator is just $\hat{\mathfrak{x}} = x$, the momentum operator is $\hat{\mathfrak{p}}_x = p_x \equiv \frac{\hbar}{i}\frac{\partial}{\partial x}$, where $\hbar = \frac{h}{2\pi}$, and the non-relativistic energy operator (hamiltonian) is $\hat{H} = \frac{\hat{\vec{\mathfrak{p}}}\hat{\vec{\mathfrak{p}}}}{2m} + \hat{V}(x) = -\frac{\hbar^2}{2m}\nabla^2 + V(x)$.

Observables are said to be *compatible* if they can be defined simultaneously with arbitrary accuracy; i.e., if they are "independent." A criterion for compatibility is the *commutator*. Two observables $\hat{A}, \hat{B}$ are compatible, if their *commutator* vanishes; i.e.,

$$\left[\hat{A}, \hat{B}\right] = \hat{A}\hat{B} - \hat{B}\hat{A} = 0 \quad . \tag{21}$$

For example, position and momentum operators[7]

$$[\hat{\mathfrak{x}}, \hat{\mathfrak{p}}_x] = \hat{\mathfrak{x}}\hat{\mathfrak{p}}_x - \hat{\mathfrak{p}}_x\hat{\mathfrak{x}} = x\frac{\hbar}{i}\frac{\partial}{\partial x} - \frac{\hbar}{i}\frac{\partial}{\partial x}x = i\hbar \neq 0 \tag{22}$$

and thus do not commute. Therefore, position and momentum of a state cannot be measured simultaneously with arbitrary accuracy. It can be shown that this property gives rise to the *Heisenberg uncertainty relations*

$$\Delta x \Delta p_x \geq \frac{\hbar}{2} \quad , \tag{23}$$

where $\Delta x$ and $\Delta p_x$ is given by $\Delta x = \sqrt{\langle x^2\rangle - \langle x\rangle^2}$ and $\Delta p_x = \sqrt{\langle p_x^2\rangle - \langle p_x\rangle^2}$, respectively.

It has recently been demonstrated that (by an analog embodiment using paricle beams) every self-adjoint operator in a finite dimensional Hilbert space can be experimentally realized [67].

*(VI)* The result of any single measurement of the observable $\hat{R}$ can only be one of the eigenvalues $r_n$ of the corresponding operator $\hat{R}$. As a result of the measurement, the

[7]the expressions should be intrepreted in the sense of operator equations; the operators themselves act on states.

system is in (one of) the state(s) $| a, r_n\rangle$ of $\hat{R}$ with the associated eigenvalue $r_n$ and not in a coherent superposition. This has been given rise to speculations concerning the "collapse of the wave function (state)." But, as has been argued recently (cf. [37]), it is possible to reconstruct coherence; i.e., to "reverse the collapse of the wave function (state)" if the process of measurement is reversible. After this reconstruction, no information about the measurement must be left, not even in principle. How did Schrödinger, the creator of wave mechanics, perceives the $\psi$-function? In his 1935 paper "Die Gegenwärtige Situation in der Quantenmechanik" ("The present situation in quantum mechanics" [70], p. 53), Schrödinger states,[8]

> *Die $\psi$-Funktion als Katalog der Erwartung:* … Sie [[die $\psi$-Funktion]] ist jetzt das Instrument zur Voraussage der Wahrscheinlichkeit von Maßzahlen. In ihr ist die jeweils erreichte Summe theoretisch begründeter Zukunftserwartung verkörpert, gleichsam wie in einem *Katalog* niedergelegt. … Bei jeder Messung ist man genötigt, der $\psi$-Funktion (=dem Voraussagenkatalog eine eigenartige, etwas plötzliche Veränderung zuzuschreiben, die von der *gefundenen Maßzahl* abhängt und sich *nicht vorhersehen läßt;* woraus allein schon deutlich ist, daß diese zweite Art von Veränderung der $\psi$-Funktion mit ihrem regelmäßigen Abrollen *zwischen* zwei Messungen nicht das mindeste zu tun hat. Die abrupte Veränderung durch die Messung … ist der interessanteste Punkt der ganzen Theorie. Es ist genau *der* Punkt, der den Bruch mit dem naiven Realismus verlangt. Aus *diesem* Grund kann man die $\psi$-Funktion *nicht* direkt an die Stelle des Modells oder des Realdings setzen. Und zwar nicht etwa weil man einem Realding oder einem Modell nicht abrupte unvorhergesehene Änderungen zumuten dürfte, sondern weil vom realistischen Standpunkt die Beobachtung ein Naturvorgang ist wie jeder andere und nicht per se eine Unterbrechung des regelmäßigen Naturlaufs hervorrufen darf.

It therefore seems not unreasonable to state that, epistemologically, quantum mechanics is more a theory of knowledge of an (intrinsic) observer rather than the platonistic physics "God knows." The wave function, i.e., the state of the physical system in a particular representation (base), is a representation of the observer's knowledge; it is a representation or name or code or index of the information or knowledge the observer has access to.

[8] *The $\psi$-function as expectation-catalog:* … In it [[the $\psi$-function]] is embodied the momentarily-attained sum of theoretically based future expectation, somewhat as laid down in a *catalog.* … For each measurement one is required to ascribe to the $\psi$-function (=the prediction catalog) a characteristic, quite sudden change, which *depends on the measurement result obtained,* and so *cannot be forseen;* from which alone it is already quite clear that this second kind of change of the $\psi$-function has nothing whatever in common with its orderly development *between* two measurements. The abrupt change [[of the $\psi$-function (=the prediction catalog)]] by measurement … is the most interesting point of the entire theory. It is precisely *the* point that demands the break with naive realism. For *this* reason one cannot put the $\psi$-function directly in place of the model or of the physical thing. And indeed not because one might never dare impute abrupt unforseen changes to a physical thing or to a model, but because in the realism point of view observation is a natural process like any other and cannot *per se* bring about an interruption of the orderly flow of natural events.

*(VII)* The *average value* or *expectation value* of an observable $\hat{R}$ in the state $\mid \psi\rangle$ is given by

$$\langle R\rangle = \langle \psi \mid R \mid \psi\rangle = \sum_{n,a} r_n \langle \psi \mid a, r_n\rangle\langle a, r_n \mid \psi\rangle = \sum_{n,a} r_n |\langle \psi \mid a, r_n\rangle|^2 \quad . \tag{24}$$

*(VIII)* The probability to find a system represented by state $\mid \psi\rangle$ in some state $\mid i\rangle$ of the orthonormalized basis is given by

$$|\langle i \mid \psi\rangle|^2 \quad . \tag{25}$$

For the continuous case, the probability of finding the system between $i$ and $i+di$ is given by

$$|\langle i \mid \psi\rangle|^2 di \quad . \tag{26}$$

*(IX)* The dynamical law or equation of motion can be written in the form

$$\mid \psi(t)\rangle = \hat{U} \mid \psi(t_0)\rangle \quad , \tag{27}$$

where $\hat{U}^\dagger = \hat{U}^{-1}$, i.e., $\hat{U}\hat{U}^\dagger = \hat{U}^\dagger\hat{U} = 1$ is a linear *unitary evolution operator*. So, all quantum dynamics is based on *linear* operations! This fact is of central importance in interferometry and, with computing being interpretable as interferometry, for the theory of quantum computability.

The *Schrödinger equation*

$$i\hbar\frac{\partial}{\partial t} \mid \psi(t)\rangle = \hat{H} \mid \psi(t)\rangle \tag{28}$$

is obtained by identifying $\hat{U}$ with

$$\hat{U} = e^{-i\hat{H}t/\hbar} \quad , \tag{29}$$

where $\hat{H}$ is a self-adjoint hamiltonian ("energy") operator, and by differentiating (27) with respect to the time variable $t$ and using (29); i.e.,

$$\frac{\partial}{\partial t} \mid \psi(t)\rangle = -\frac{i\hat{H}}{\hbar}e^{-i\hat{H}t/\hbar} \mid \psi(t_0)\rangle = -\frac{i\hat{H}}{\hbar} \mid \psi(t)\rangle \quad . \tag{30}$$

In terms of the set orthonormal base vectors $\{\mid i\rangle \mid i \in \mathbb{I}\}$, the Schrödinger equation (28) can be written as

$$i\hbar\frac{\partial}{\partial t}\langle i \mid \psi(t)\rangle = \sum_{j\in\mathbb{I}}\langle i \mid H \mid j\rangle\langle j \mid \psi(t)\rangle \quad ,33 \tag{31}$$

with $\langle i \mid H \mid j\rangle = H_{ij}$ for a finite dimensional Hilbert space. Again, the sums become integrals and $\langle i \mid H \mid j\rangle = H(i, j)$ for continuous spectra. In the case of position base states $\psi(x, t) = \langle x \mid \psi(t)\rangle$, the Schrödinger equation (28) takes on the form

$$i\hbar\frac{\partial}{\partial t}\psi(x, t) = \hat{H}\psi(x, t) = \left[\frac{\hat{\mathfrak{p}}\hat{\mathfrak{p}}}{2m} + \hat{V}(x)\right] \psi(x, t) = \left[-\frac{\hbar^2}{2m}\nabla^2 + V(x)\right] \psi(x, t) \quad . \tag{32}$$

*(X)* For stationary $\mid \psi_n(t)\rangle = e^{-(i/\hbar)E_n t} \mid \psi_n\rangle$, the Schrödinger equation (28) can be brought into its time-independent form

$$\hat{H} \mid \psi_n\rangle = E_n \mid \psi_n\rangle \quad . \tag{33}$$

Here, $i\hbar\frac{\partial}{\partial t} \mid \psi_n(t)\rangle = E_n \mid \psi_n(t)\rangle$ has been used; $E_n$ and $\mid \psi_n\rangle$ stand for the $n$'th eigenvalue and eigenstate of $\hat{H}$, respectively.

Usually, a physical problem is defined by the hamiltonian $\hat{H}$. The problem of finding the physically relevant states reduces to finding a complete set of eigenvalues and eigenstates of $\hat{H}$. Most elegant solutions utilize the symmetries of the problem, i.e., of $\hat{H}$. There exist two "canonical" examples, the $1/r$-potential and the harmonic oscillator potential, which can be solved wonderfully by this methods (and they are presented over and over again in standard courses of quantum mechanics), but not many more. (See [23] for a detailed treatment of various hamiltonians $\hat{H}$.)

Having now set the stage of the quantum formalism, an elementary twodimensional example of a two-state system shall be exhibited ([32], p. 8-11). Let us denote the two base states by $\mid 1\rangle$ and $\mid 2\rangle$. Any arbitrary physical state $\mid \psi\rangle$ is a coherent superposition of $\mid 1\rangle$ and $\mid 2\rangle$ and can be written as $\mid \psi\rangle = \mid 1\rangle\langle 1 \mid \psi\rangle + \mid 2\rangle\langle 2 \mid \psi\rangle$ with the two coefficients $\langle 1 \mid \psi\rangle, \langle 2 \mid \psi\rangle \in \mathbb{C}$.

Let us discuss two particular types of evolutions.

First, let us discuss the Schrödinger equation (28) with diagonal Hamilton matrix, i.e., with vanishing off-diagonal elements,

$$\langle i \mid H \mid j\rangle = \begin{pmatrix} E_1 & 0 \\ 0 & E_2 \end{pmatrix} \quad . \tag{34}$$

In this case, the Schrödinger equation (33) decouples and reduces to

$$i\hbar\frac{\partial}{\partial t}\langle 1 \mid \psi(t)\rangle = E_1\langle 1 \mid \psi(t)\rangle \quad , \qquad i\hbar\frac{\partial}{\partial t}\langle 2 \mid \psi(t)\rangle = E_2\langle 2 \mid \psi(t)\rangle \quad , \tag{35}$$

resulting in

$$\langle 1 \mid \psi(t)\rangle = ae^{-iE_1 t/\hbar} \quad , \qquad \langle 2 \mid \psi(t)\rangle = be^{-iE_2 t/\hbar} \quad , \tag{36}$$

with $a, b \in \mathbb{C}$, $|a|^2 + |b|^2 = 1$. These solutions correspond to *stationary states* which do not change in time; i.e., the probability to find the system in the two states is constant

$$|\langle 1 \mid \psi\rangle|^2 = |a|^2 \quad , \qquad |\langle 2 \mid \psi\rangle|^2 = |b|^2 \quad . \tag{37}$$

Second, let us discuss the Schrödinger equation (33) with with non-vanishing but equal off-diagonal elements $-A$ and with equal diagonal elements $E$ of the hamiltonian matrix; i.e.,

$$\langle i \mid H \mid j\rangle = \begin{pmatrix} E & -A \\ -A & E \end{pmatrix} \quad . \tag{38}$$

In this case, the Schrödinger equation (33) reads

$$i\hbar\frac{\partial}{\partial t}\langle 1 \mid \psi(t)\rangle = E\langle 1 \mid \psi(t)\rangle - A\langle 2 \mid \psi(t)\rangle \quad , \tag{39}$$

$$i\hbar\frac{\partial}{\partial t}\langle 2 \mid \psi(t)\rangle = E\langle 2 \mid \psi(t)\rangle - A\langle 1 \mid \psi(t)\rangle \quad . \tag{40}$$

These equations can be solved in a number of ways. For example, taking the sum and the difference of the two, one obtains

$$i\hbar\frac{\partial}{\partial t}(\langle 1 \mid \psi(t)\rangle + \langle 2 \mid \psi(t)\rangle) = (E-A)(\langle 1 \mid \psi(t)\rangle + \langle 2 \mid \psi(t)\rangle) \quad , \tag{41}$$

$$i\hbar\frac{\partial}{\partial t}(\langle 1 \mid \psi(t)\rangle - \langle 2 \mid \psi(t)\rangle) = (E+A)(\langle 1 \mid \psi(t)\rangle - \langle 2 \mid \psi(t)\rangle) \quad . \tag{42}$$

The solution are again two stationary states

$$\langle 1 \mid \psi(t)\rangle + \langle 2 \mid \psi(t)\rangle = ae^{-(i/h)(E-A)t} \quad , \tag{43}$$

$$\langle 1 \mid \psi(t)\rangle - \langle 2 \mid \psi(t)\rangle = be^{-(i/h)(E+A)t} \quad . \tag{44}$$

Thus,

$$\langle 1 \mid \psi(t)\rangle = \frac{a}{2}e^{-(i/h)(E-A)t} + \frac{b}{2}e^{-(i/h)(E+A)t} \quad , \tag{45}$$

$$\langle 2 \mid \psi(t)\rangle = \frac{a}{2}e^{-(i/h)(E-A)t} - \frac{b}{2}e^{-(i/h)(E+A)t} \quad . \tag{46}$$

Assume now that initially, i.e., at $t = 0$, the system was in state $\mid 1\rangle =\mid \psi(t = 0)\rangle$. This assumption corresponds to $\langle 1 \mid \psi(t = 0)\rangle = 1$ and $\langle 2 \mid \psi(t = 0)\rangle = 0$. What is the probability that the system will be found in the state $\mid 2\rangle$ at the time $t > 0$, or that it will still be found in the state $\mid 1\rangle$ at the time $t > 0$? Setting $t = 0$ in equations (45) and (46) yields

$$\langle 1 \mid \psi(t = 0)\rangle = \frac{a+b}{2} = 1 \quad , \qquad \langle 2 \mid \psi(t = 0)\rangle = \frac{a-b}{2} = 0 \quad , \tag{47}$$

and thus $a = b = 1$. Equations (45) and (46) can now be evaluated at $t > 0$ by substituting 1 for $a$ and $b$,

$$\langle 1 \mid \psi(t)\rangle = e^{-(i/h)Et}\left[\frac{e^{(i/h)At} + e^{-(i/h)At}}{2}\right] = e^{-(i/h)Et}\cos\frac{At}{\hbar} \quad , \tag{48}$$

$$\langle 2 \mid \psi(t)\rangle = e^{-(i/h)Et}\left[\frac{e^{(i/h)At} - e^{-(i/h)At}}{2}\right] = i\,e^{-(i/h)Et}\sin\frac{At}{\hbar} \quad . \tag{49}$$

Finally, the probability that the system is in state $\mid 1\rangle$ and $\mid 2\rangle$ is

$$|\langle 1 \mid \psi(t)\rangle|^2 = \cos^2\frac{At}{\hbar} \quad , \qquad |\langle 2 \mid \psi(t)\rangle|^2 = \sin^2\frac{At}{\hbar} \quad , \tag{50}$$

respectively. This results in a constant transition probability back and forth a given state, as depicted in Fig. 3.

Let us shortly mention one particular realization of a two-state system which, among many others, has been discussed in the Feynman lectures [32]. Consider an ammonia ($NH_3$) molecule. If one fixes the plane spanned by the three hydrogen atoms, one observes two possible spatial configurations $\mid 1\rangle$ and $\mid 2\rangle$, corresponding to position of the nitrogen atom in the lower or the upper hemisphere, respectively (cf. Fig. 4). The nondiagonal elements of the hamiltonian $H_{12} = H_{21} = -A$ correspond to a nonvanishing transition probability from one such configuration into the other. If the ammonia has been originally in state $\mid 1\rangle$, it will constantly swing back and forth between the two states, with a probability given by equations (50).

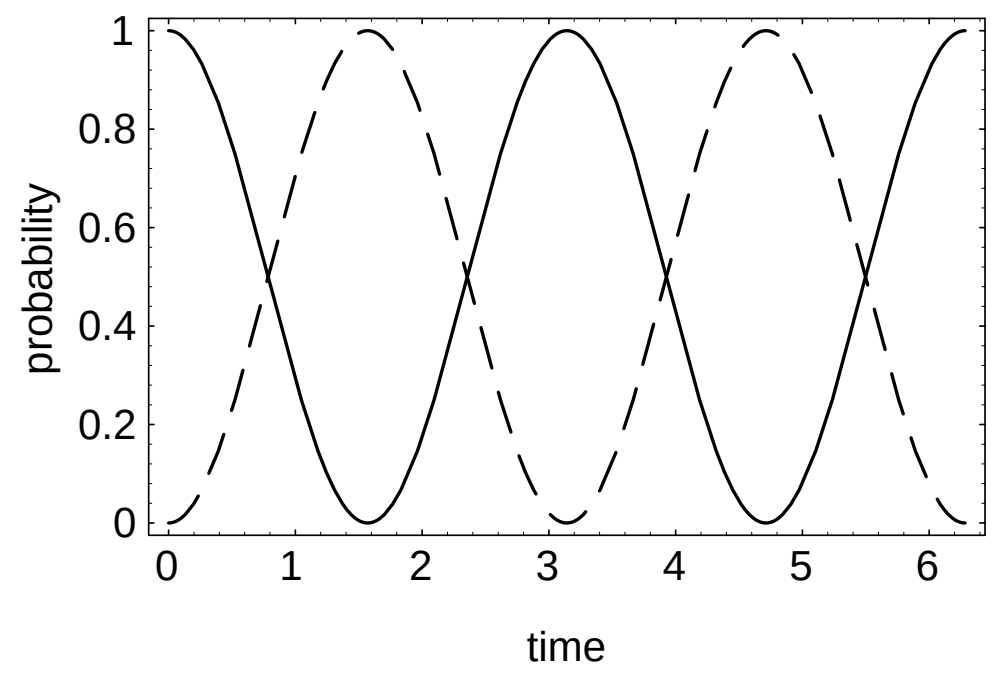

Figure 3: The probabilities $|\langle 1 \mid \psi(t)\rangle|^2 = \cos^2(At/\hbar)$ (solid line) and $|\langle 2 \mid \psi(t)\rangle|^2 = \sin^2(At/\hbar)$ (dashed line) as a function of time (in units of $\hbar/A$) for a quantized system which is in state $|\,1\rangle$ at $t = 0$.

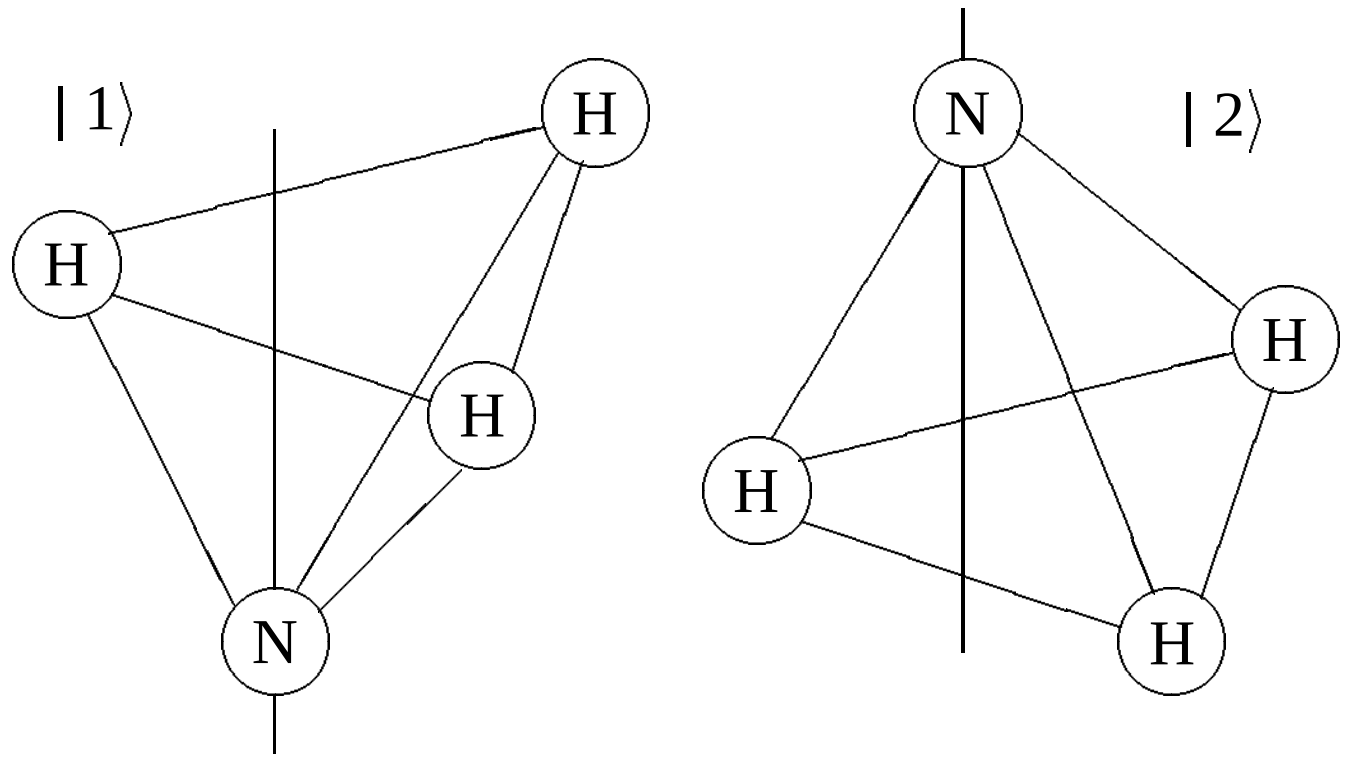

Figure 4: The two equivalent geometric arrangements of the ammonia ($NH_3$) molecule.

## 2.2 From single to multiple quanta — "second" field quantization

The quantum formalism developed so far is about *single* quantized objects. What if one wants to consider many such objects? Do we have to add assumptions in order to treat such multi-particle, multi-quanta systems appropriately?

The answer is yes. Experiment and theoretical reasoning (the representation theory of the Lorentz group [75] and the spin-statistics theorem [44, 53, 11, 42]) indicate that there are (at least) two basic types of states (quanta, particles): *bosonic* and *fermionic* states. Bosonic states have what is called "integer spin;" i.e., $s_b = 0, \hbar, 2\hbar, 3\hbar, \ldots$, whereas fermionic states have "half-integer spin;" $s_f = \frac{1\hbar}{2}, \frac{3\hbar}{2}, \frac{5\hbar}{2} \ldots$. Most important, they are characterized by the way identical copies of them can be "brought together." Consider two boxes, one for identical bosons, say photons, the other one for identical fermions, say electrons. For the first, bosonic box, the probability that another identical boson is added *increases with the number of identical bosons* which are already in the box. There is a tendency of bosons to "condensate" into the same state. The second, fermionic box, behaves quite differently. If it is already occupied by one fermion, another identical fermion cannot enter. This is expressed in the *Pauli exclusion principle:* A system of fermions can never occupy a configuration of individual states in which two individual states are identical.

How can the bose condensation and the Pauli exclusion principle be implemented? There are several forms of implementation (e.g., fermionic behavior via Slater-determinants), but the most compact and widely practiced form uses operator algebra. In the following we shall present this formalism in the context of quantum field theory [40, 51, 44, 53, 11, 42, 35].

A *classical* field can be represented by its Fourier transform ("$*$" stands for complex conjugation)

$$
\begin{aligned}
A(x,t) &= A^{(+)}(x,t) + A^{(-)}(x,t) && (51)\\
A^{(+)}(x,t) &= [A^{(-)}(x,t)]^{*} && (52)\\
A^{(+)}(x,t) &= \sum_{k_i,s_i} a_{k_i,s_i} u_{k_i,s_i}(x) e^{-i\omega_{k_i} t} \quad , && (53)
\end{aligned}
$$

where $\nu = \omega_{k_i}/2\pi$ stands for the frequency in the field mode labelled by momentum $k_i$ and $s_i$ is some observable such as spin or polarization. $u_{k_i,s_i}$ stands for the polarization vector (spinor) at $k_i, s_i$, and, most important with regards to the quantized case, *complex-valued* Fourier coefficients $a_{k_i,s_i} \in \mathbb{C}$.

From now on, the $k_i, s_i$-mode will be abbreviated by the symbol $i$; i.e., $1 \equiv k_1, s_1$, $2 \equiv k_2, s_2$, $3 \equiv k_3, s_3$, $\ldots$, $i \equiv k_i, s_i$, $\ldots$.

In (second[9]) quantization, the classical Fourier coefficients $a_i$ become re-interpreted as *operators*, which obey the following algebraic rules (scalars would not do the trick). For *bosonic* fields (e.g., for the electromagnetic field), the *commutator* relations are ("†" stands for self-adjointness):

$$
\left[a_i, a_j^\dagger\right] = a_i a_j^\dagger - a_j^\dagger a_i = \delta_{ij} \quad , \qquad (54)
$$

[9]of course, there is only "the one and only" quantization, the term "second" often refers to operator techniques for multiqanta systems; i.e., quantum field theory

$$[a_i, a_j] = \left[a_i^\dagger, a_j^\dagger\right] = 0 \quad . \tag{55}$$

For *fermionic* fields (e.g., for the electron field), the *anti-commutator* relations are:

$$\{a_i, a_j^\dagger\} = a_i a_j^\dagger + a_j^\dagger a_i = \delta_{ij} \quad , \tag{56}$$

$$\{a_i, a_j\} = \{a_i^\dagger, a_j^\dagger\} = 0 \quad . \tag{57}$$

The anti-commutator relations, in particular $\{a_j^\dagger, a_j^\dagger\} = 2(a_j^\dagger)^2 = 0$, are just a formal expression of the Pauli exclusion principle stating that, unlike bosons, two or more identical fermions cannot co-exist.

The operators $a_i^\dagger$ and $a_i$ are called *creation* and *annihilation* operators, respectively. This terminology suggests itself if one introduces *Fock states* and the *occupation number formalism*. $a_i^\dagger$ and $a_i$ are applied to Fock states to following effect.

The Fock space associated with a quantized field will be the direct product of all Hilbert spaces $\mathfrak{H}_i$; i.e.,

$$\prod_{i \in \mathbb{I}} \mathfrak{H}_i \quad , \tag{58}$$

where $\mathbb{I}$ is an index set characterizing all different field modes labeled by $i$. Each boson (photon) field mode is equivalent to a harmonic oscillator [35, 52]; each fermion (electron, proton, neutron) field mode is equivalent to the Larmor precession of an electron spin.

In what follows, only finite-size systems are studied. The Fock states are based upon the Fock vacuum. The Fock vacuum is a direct product of states $| 0_i\rangle$ of the $i$'th Hilbert space $\mathfrak{H}_i$ characterizing mode $i$; i.e.,

$$\begin{aligned} | 0\rangle &= \prod_{i \in \mathbb{I}} | 0\rangle_i = | 0\rangle_1 \otimes | 0\rangle_2 \otimes | 0\rangle_3 \otimes \cdots \\ &= | \bigcup_{i \in \mathbb{I}} \{0_i\}\rangle = | \{0_1, 0_2, 0_3, \ldots\}\rangle \quad , \end{aligned} \tag{59}$$

where again $\mathbb{I}$ is an index set characterizing all different field modes labeled by $i$. "$0_i$" stands for 0 (no) quantum (particle) in the state characterized by the quantum numbers $i$. Likewise, more generally, "$N_i$" stands for $N$ quanta (particles) in the state characterized by the quantum numbers $i$.

The annihilation operators $a_i$ are designed to destroy one quantum (particle) in state $i$:

$$a_j | 0\rangle = 0 \quad , \tag{60}$$

$$\begin{aligned} & a_j | \{0_1, 0_2, 0_3, \ldots, 0_{j-1}, N_j, 0_{j+1}, \ldots\}\rangle = \\ & \quad = \sqrt{N_j} | \{0_1, 0_2, 0_3, \ldots, 0_{j-1}, (N_j - 1), 0_{j+1}, \ldots\}\rangle \quad . \end{aligned} \tag{61}$$

The creation operators $a_i^\dagger$ are designed to create one quantum (particle) in state $i$:

$$a_j^\dagger | 0\rangle = | \{0_1, 0_2, 0_3, \ldots, 0_{j-1}, 1_j, 0_{j+1}, \ldots\}\rangle \quad . \tag{62}$$

More generally, $N_j$ operators $(a_j^\dagger)^{N_j}$ create an $N_j$-quanta (particles) state

$$(a_j^\dagger)^{N_j} | 0\rangle \propto | \{0_1, 0_2, 0_3, \ldots, 0_{j-1}, N_j, 0_{j+1}, \ldots\}\rangle \quad . \tag{63}$$

For fermions, $N_j \in \{0,1\}$ because of the Pauli exclusion principle. For bosons, $N_j \in \mathbb{N}_0$. With proper normalization [which can motivated by the (anti-)commutator relations and by $|\langle X \mid X\rangle|^2 = 1$], a state containing $N_1$ quanta (particles) in mode 1, $N_2$ quanta (particles) in mode 2, $N_3$ quanta (particles) in mode 3, *etc.*, can be generated from the Fock vacuum by

$$\mid \bigcup_{i\in\mathbb{I}}\{N_i\}\rangle \equiv \mid \{N_1, N_2, N_3, \ldots\}\rangle = \prod_{i\in\mathbb{I}} \frac{(a_i^\dagger)^{N_i}}{\sqrt{N_i!}} \mid 0\rangle \quad . \tag{64}$$

The most general quantized field configuration in the Fock basis $\mid X\rangle$ is thus a coherent superposition of such quantum states (64) with weights $f_{\{N_i\}} \in \mathbb{C}$; i.e.,

$$\mid X\rangle = \sum_{\bigcup_{i\in\mathbb{I}}\{N_i\}} f_{\bigcup_{i\in\mathbb{I}}\{N_i\}} \mid \bigcup_{i\in\mathbb{I}}\{N_i\}\rangle \quad . \tag{65}$$

Compare (65) to the classical expression (53). Classically, the most precise specification has been achieved by specifying one complex number $a_i \in \mathbb{C}$ for every field mode $i$. Quantum mechanically, we have to sum over a "much larger" set $\bigcup_{i\in\mathbb{I}}\{N_i\} \in \{\{0_1, 0_2, 0_3, \ldots\}, \{1_1, 0_2, 0_3, \ldots\}, \{0_1, 1_2, 0_3, \ldots\}, \{0_1, 0_2, 1_3, \ldots\}, \ldots \{1_1, 1_2, 0_3, \ldots\}, \ldots\}$, which results from additional (nonclassical) opportunities to occupy every boson field mode with $0, 1, 2, 3, \ldots$ quanta (particles).

Even if the field would consist of only one mode $k, s$, for bosons, there is a countable infinite ($\aleph_0$) set of complex coefficients $\{f_0, f_1, f_2, f_3, \ldots\}$ in the field specification. (For fermions, only two coefficient $\{f_0, f_1\}$ would be required, corresponding to a nonfilled and a filled mode.) For such a bosonic one-mode field, the summation in (65) reduces to

$$\mid X\rangle = \sum_{N=0}^{\infty} f_N \mid N\rangle \quad , \tag{66}$$

with the normalization condition

$$|\langle X \mid X\rangle|^2 = \sum_{N=0}^{\infty} |f_N|^2 = 1 \quad . \tag{67}$$

Thus, as has been stated by Glauber ([35], p. 64),

> … in quantum theory, there is an infinite set of complex numbers which specifies the state of a single mode. This is in contrast to classical theory where each mode may be described by a single complex number. This shows that there is vastly more freedom in quantum theory to invent states of the world than there is in the classical theory. We cannot think of quantum theory and classical theory in one-to-one terms at all. In quantum theory, there exist whole spaces which have no classical analogues, whatever.

## 2.3 Quantum interference

In what follows a few quantum interference devices will be reviewed. Thereby, we shall make use of a simple "toolbox"-scheme of combining lossless elements of an experimental setup for the theoretical calculation [38]. The elements of this "toolbox" are listed in

$$\mid i\rangle \mid j\rangle \equiv a_i^\dagger a_j^\dagger \mid 0\rangle = \mid i\rangle \otimes \mid j\rangle = \mid 0_1, 0_2, 0_3, \ldots, 0_{i-1}, 1_i, 0_{i+1}, \ldots, 0_{j-1}, 1_j, 0_{j+1}, \ldots\rangle \quad . \tag{68}$$

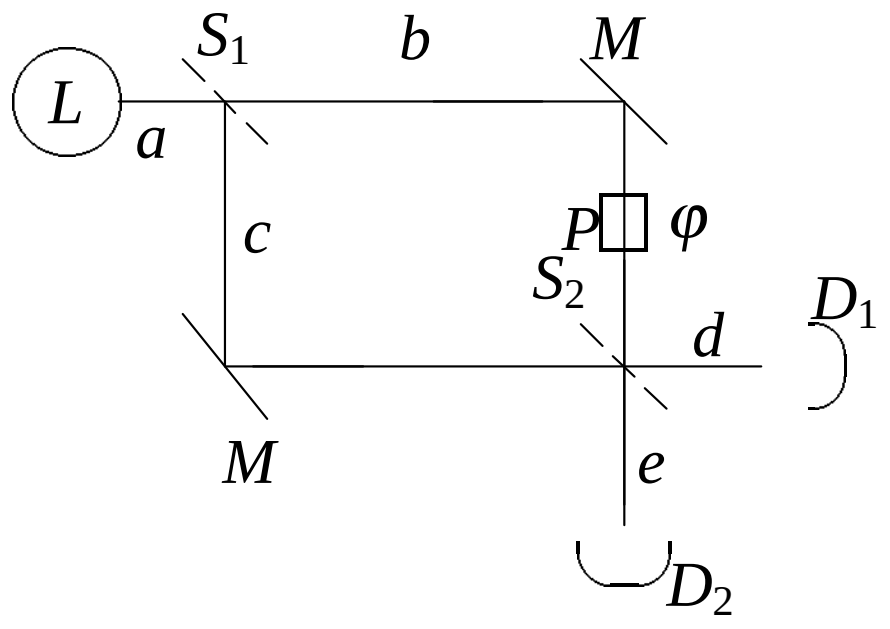

Figure 5: Mach-Zehnder interferometer. A single quantum (photon, neutron, electron *etc*) is emitted in $L$ and meets a lossless beam splitter (half-silvered mirror) $S_1$, after which its wave function is in a coherent superposition of $\mid b\rangle$ and $\mid c\rangle$. In beam path $b$ a phase shifter shifts the phase of state $\mid b\rangle$ by $\varphi$. The two beams are then recombined at a second lossless beam splitter (half-sivered mirror) $S_2$. The quant is detected at either $D_1$ or $D_2$, corresponding to the states $\mid d\rangle$ and $\mid e\rangle$, respectively.

Table 1. These "toolbox" rules can be rigorously motivated by the full quantum optical calculations (e.g., [86, 13]) but are much easier to use. Note that, in the notation used, for $i < j$,

In present-day quantum optical nonlinear devices (NL), parametric up- or down-conversion, i.e., the production of a single quant (particle) from two field quanta (particles) and the production of two field quanta (particles) from a single one occurs at the very low amplitude rate of $\eta \approx 10^{-6}$. $T$ and $R = \sqrt{1-T^2}$ are transmission and reflection coefficients. Notice that the two-slit device is not elementary: it can be realized by a beam splitter (half-silvered mirror) and a successive phase shift of $\varphi = -\pi$ in the reflected channel; i.e., $|a\rangle \to (|b\rangle + i|c\rangle)/\sqrt{2} \to (|b\rangle + ie^{-i\pi/2}|c\rangle)/\sqrt{2} \to (|b\rangle|c\rangle)/\sqrt{2}$.

Let us start with a *Mach-Zehnder* interferometer drawn in Fig. 5. The computation proceeds by successive substitution (transition) of states; i.e.,

$$S_1 : \; |a\rangle \;\to\; (|b\rangle + i|c\rangle)/\sqrt{2} \quad , \tag{69}$$

$$P : \; |b\rangle \;\to\; |b\rangle e^{i\varphi} \quad , \tag{70}$$

$$S_2 : \; |b\rangle \;\to\; (|e\rangle + i|d\rangle)/\sqrt{2} \quad , \tag{71}$$

$$S_2 : \; |c\rangle \;\to\; (|d\rangle + i|e\rangle)/\sqrt{2} \quad . \tag{72}$$

The resulting transition is.[10]

$$\mid a\rangle \to \mid \psi\rangle = i\left(\frac{e^{i\varphi}+1}{2}\right)\mid d\rangle + \left(\frac{e^{i\varphi}-1}{2}\right)\mid e\rangle \quad . \tag{73}$$

Assume that $\varphi = 0$, i.e., there is no phase shift at all. Then, equation (73) reduces to $\mid a\rangle \to i \mid d\rangle$, and the emitted quant is detected only by $D_1$. Assume that $\varphi = \pi$. Then, equation (73) reduces to $\mid a\rangle \rightarrow -\mid e\rangle$, and the emitted quant is detected only by $D_2$. If one varies the phase shift $\varphi$, one obtains the following detection probabilities, which are identical to the probabilities drawn in Fig. 3 with the substitutions $1 \rightarrow d$, $2 \rightarrow e$, $2At/\hbar \rightarrow \varphi$ and time $\rightarrow$ phase shift.

[10] A *Mathematica* program for this computation is in appendix C.1.

| physical process | symbol | state transformation |
|---|---|---|
| reflection by mirror | 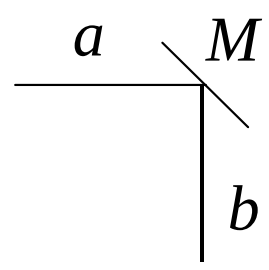  | $\|a\rangle \to \|b\rangle = \|a\rangle$ |
| beam splitter | 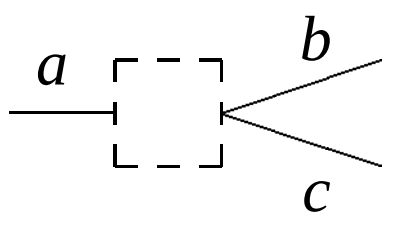  | $\|a\rangle \to (\|b\rangle + \|c\rangle)/\sqrt{2}$ |
| transmission/reflection<br>by a beam splitter<br>(half-silvered mirror) | 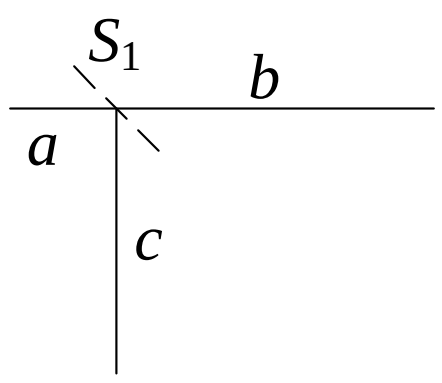  | $\|a\rangle \to (\|b\rangle + i\|c\rangle)/\sqrt{2}$<br>$\|a\rangle \to T\|b\rangle + iR\|c\rangle,$<br>$T^2 + R^2 = 1,\ T, R \in [0,1]$ |
| phase-shift $\varphi$ | 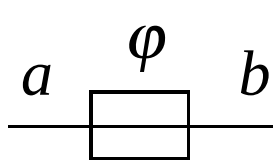  | $\|a\rangle \to \|b\rangle = \|a\rangle e^{i\varphi}$ |
| parametric down-conversion | 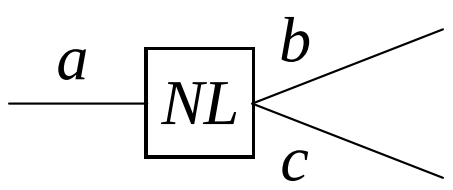  | $\|a\rangle \to \eta\|b\rangle\|c\rangle$ |
| parametric up-conversion | 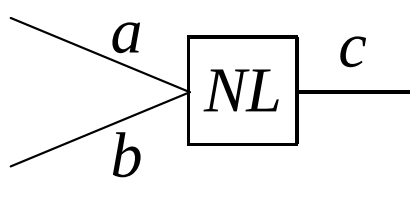  | $\|a\rangle \mid b\rangle \to \eta\|c\rangle$ |
| amplification | 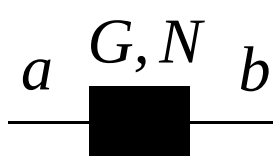  | $\|A_i\rangle\|a\rangle \to \|b; G, N\rangle$ |

Table 1: "Toolbox"lossless elements for quantum interference devices.

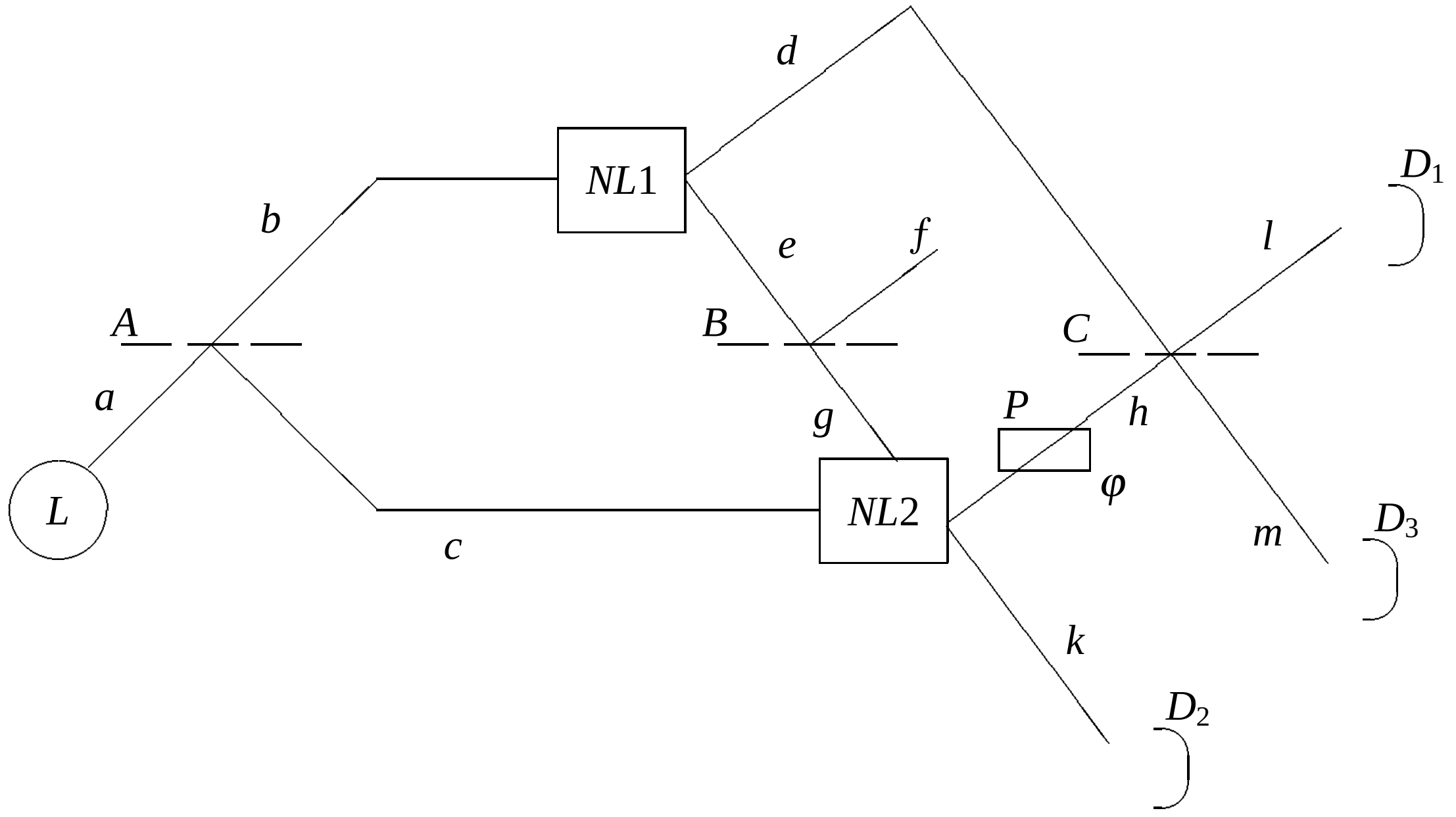

Figure 6: Mandel interferometer. Manipulating one photon can alter the interference pattern of another. The arrangement can produce an interference pattern at detector $D_1$ when the phase shifter $P$ is varied. An entering ultraviolet photon $a$ is split at beam splitter (half-silvered mirror) $A$ so that both down-conversion cryststals $NL_1$ and $NL_2$ are illuminated. One of the resulting pair of downconversion photons can reach $D_1$ by way of beam path $d$ or $h$ If one could monitor beams $e$ and $k$ separately, one would know in which crystal the down-conversion occurred, and there would be no interference. But merging beams $e$ and $k$ in this configuration lets the alternative paths of the other photon interfere.

$$P_{D_1}(\varphi) = |\langle d \mid \psi\rangle|^2 = \cos^2(\frac{\varphi}{2}) \quad , \quad P_{D_2}(\varphi) = |\langle e \mid \psi\rangle|^2 = \sin^2(\frac{\varphi}{2}) \quad . \tag{74}$$

For some “mindboggling” features of Mach-Zehnder interfereometry, see [7].

So far, only a single quantum (particle) at a time was involved. Could we do two-particle or multiparticle interferometry?

Fig. 6 shows an arrangement in which manipulation of one quantum (photon) can alter the interference pattern of another quantum (photon) [55, 38]. The computation of the process again proceeds by successive substitution (transition) of states; i.e.,

$$\begin{aligned}
A: \ |a\rangle &\rightarrow (|b\rangle + i|c\rangle)/\sqrt{2} \quad , &(75)\\
NL_1: \ |b\rangle &\rightarrow \eta|d\rangle|e\rangle \quad , &(76)\\
NL_2: \ |c\rangle &\rightarrow \eta|h\rangle|k\rangle \quad , &(77)\\
P: \ |h\rangle &\rightarrow |h\rangle e^{i\varphi} \quad , &(78)\\
B: \ |e\rangle &\rightarrow T|g\rangle + iR|f\rangle \quad , &(79)\\
C: \ |h\rangle &\rightarrow (|l\rangle + i|m\rangle)/\sqrt{2} \quad , &(80)\\
C: \ |d\rangle &\rightarrow (|m\rangle + i|l\rangle)/\sqrt{2} \quad , &(81)\\
|g\rangle &\rightarrow |k\rangle \quad . &(82)
\end{aligned}$$

The resulting transition is.[11]

$$|a\rangle \rightarrow |\psi\rangle = \frac{\eta}{2}\left\{i\left(e^{i\varphi}+T\right)|k\rangle|l\rangle - e^{i\varphi}|k\rangle|m\rangle - R|f\rangle|l\rangle + iR|f\rangle|m\rangle + T|k\rangle|m\rangle\right\} \tag{83}$$

Let us first consider only those two-photon events which occur simultaneously at detectors $D_1$ and $D_2$; i.e., we are interested in the state $|l\rangle|k\rangle$. The probability for such events is given by

$$|\langle l|\langle k|\psi\rangle|^2 = \frac{\eta^2}{4}\left(1+T^2+2T\cos\varphi\right) \quad . \tag{84}$$

Let us now consider only those two-photon events which occur simultaneously at detectors $D_1$ and $D_3$; i.e., we are interested in the state $|l\rangle|f\rangle$. With the assumption that $R=\sqrt{1-T^2}$, the probability for such events is given by

$$|\langle l|\langle f|\psi\rangle|^2 = \frac{\eta^2}{4}R^2 = \frac{\eta^2}{4}\left(1-T^2\right) \quad . \tag{85}$$

Both probabilities (84) and (85) combined yield the probability to detect *any single* photon at all in detector $D_1$. It is given by

$$|\langle l|\psi\rangle|^2 = |\langle l|\langle k|\psi\rangle|^2 + |\langle l|\langle f|\psi\rangle|^2 = \frac{\eta^2}{4}\left(1+T\cos\varphi\right) \quad . \tag{86}$$

The “mindboggling” feature of the setup, as revealed by (86), is the fact that the particle detected in $D_1$ shows an interference pattern *although it did not path the phase shifter P*! The ultimate reason for this (which can be readily verified by varying $T$) is that it is impossible for detector $D_2$ to discriminate between beam path $k$ (second particle) and beam path $g$ (first photon). By this impossibility to know, the two particles become “entangled” (“Verschränkung”, a word created by Schrödinger [70]).

Is it possible to use the Mandel interferometer to communicate faster-than-light; e.g., by observing changes of the probability to detect the first particle in $D_1$ corresponding to variations of the phase shift $\varphi$ (at spatially separated points) in the path of the second particle? No, because in order to maintain coherence, i.e., in order not to be able to distinguish between the two particles in $k$ and thus to make the crucial substitution

[11] A *Mathematica* program for this computation is in appendix C.1.

$|g\rangle \to |k\rangle$, the arrangement cannot be arbitrarily spatially extended. The consistency or "peaceful coexistince" [72, 73] between relativity theory and quantum mechanics, this second "mindboggling" feature of quantum mechanics, seems to be not invalidated so far [41, 85, 54, 60, 36, 14].

## 2.4 Hilbert lattices and quantum logic

G. Birkhoff and J. von Neumann suggested [10], that, roughly speaking, the "logic of quantum events" — or, by another wording, *quantum logic* or the *quantum propositional calculus* — should be obtainable from the formal representation of physical properties.

Since, in this formalism, projection operators correspond to the physical properties of a quantum system, quantum logics is modelled in order to be isomorphic to the lattice of projections $\mathfrak{P}(\mathfrak{H})$ of the Hilbert space $\mathfrak{H}$, which in turn is isomorphic to the lattice $\mathfrak{C}(\mathfrak{H})$ of the set of subspaces of a Hilbert space. I.e., by assuming the physical validity of the quantum Hilbert space formalism, the corresponding isomorphic logical structure is investigated.

In this approach, quantum theory comes first and the logical structure of the phenomena are derived by analysing the theory, this could be considered a *"top-down"* method.

The projections $P_n$ correspond to the physical properties of a quantum system and stands for a yes/no-proposition. In J. von Neumann's words ([83], English translation, p. 249),

> *Apart from the physical quantities $\mathfrak{R}$, there exists another category of concepts that are important objects of physics — namely the properties of the states of the system S. Some such properties are: that a certain quantity $\mathfrak{R}$ takes the value $\lambda$ — or that the value of $\mathfrak{R}$ is positive —* $\cdots$
>
> *To each property $\mathfrak{E}$ we can assign a quantity which we define as follows: each measurement which distinguishes between the presence or absence of $\mathfrak{E}$ is considered as a measurement of this quantity, such that its value is* $1$ *if $\mathfrak{E}$ is verified, and zero in the opposite case. This quantity which corresponds to $\mathfrak{E}$ will also be denoted by $\mathfrak{E}$.*
>
> *Such quantities take only the values of* $0$ *and* $1$*, and conversely, each quantity $\mathfrak{R}$ which is capable of these two values only, corresponds to a property $\mathfrak{E}$ which is evidently this: "the value of $\mathfrak{R}$ is* $\neq 0$*." The quantities $\mathfrak{E}$ that correspond to the properties are therefore characterized by this behavior.*
>
> *That $\mathfrak{E}$ takes on only the values* $0, 1$ *can also be formulated as follows: Substituting $\mathfrak{E}$ into the polynomial $F(\lambda) = \lambda - \lambda^2$ makes it vanish identically. If $\mathfrak{E}$ has the operator $E$, then $F(\mathfrak{E})$ has the operator $F(E) = E - E^2$, i.e., the condition is that $E - E^2 = 0$ or $E = E^2$. In other words: the operator $E$ of $\mathfrak{E}$ is a projection.*
>
> *The projections $E$ therefore correspond to the properties $\mathfrak{E}$ (through the agency of the corresponding quantities $\mathfrak{E}$ which we just defined). If we introduce, along with the projections $E$, the closed linear manifold $\mathfrak{M}$, belonging*

| quantum logic | sign | Hilbert space entity | sign |
|---|---|---|---|
| elementary yes-no proposition | $a$ | linear subspace | $v(a)$ |
| falsity | $\mathbf{0}$ | 0-dimensional subspace | $v(0)$ |
| tautology | $\mathbf{1}$ | entire Hilbert space | $\mathfrak{H}$ |
| lattice operation | sign | Hilbert space operation | sign |
| order relation | $\preceq$ | subspace relation | $\subset$ |
| "meet" | $\sqcap$ | intersection of subspaces | $\cap$ |
| "join" | $\sqcup$ | closure of subspace spanned by subspaces | $\oplus$ |
| "orthocomplement" | $'$ | orthogonal subspace | $\perp$ |

Table 2: Identification of quantum logical entities with objects of Hilbert lattices.

> *to them ($E = P_{\mathfrak{M}}$), then the closed linear manifolds correspond equally to the properties of $\mathfrak{C}$.*

More precisely, consider the *Hilbert lattice* $\mathfrak{C}(\mathfrak{H}) = \langle B, \mathbf{0}, \mathbf{1}, \neg, \sqcup, \sqcap \rangle$ of an $n$-dimensional Hilbert space $\mathfrak{H}$, with

***(i)*** $B$ is the set of linear subspaces of $\mathfrak{H}$;

***(ii)*** $\mathbf{0}$ is the 0-dimensional subspace, $\mathbf{1}$ is the entire Hilbert space $\mathfrak{H}$;

***(iii)*** $\neg a$ is the orthogonal complement of $a$; and

***(iv)*** $a \sqcup b$ is the closure of the linear span of $a$ and $b$; and

***(v)*** $a \sqcap b$ is the intersection of $a$ and $b$.

The identification of elements, relations and operations in lattice theory with relations and operations in Hilbert space is represented in table 2.

$\mathfrak{C}(\mathfrak{H})$ is an orthocomplemented lattice. In general, $\mathfrak{C}(\mathfrak{H})$ is not distributive. Therefore, *classical (Boolean) propositional calculus is not valid* for microphysics! Let, for instance, $\mathfrak{S}', \mathfrak{S}, \mathfrak{S}^\perp$ be subsets of a Hilbert space $\mathfrak{H}$ with $\mathfrak{S}' \neq \mathfrak{S}$, $\mathfrak{S}' \neq \mathfrak{S}^\perp$, then (see Fig. 7, drawn from J. M. Jauch [45], p. 27)

$$\mathfrak{S}' \sqcap (\mathfrak{S} \sqcup \mathfrak{S}^\perp) = \mathfrak{S}' \sqcap \mathfrak{H} = \mathfrak{S}', \text{ whereas} \tag{87}$$

$$(\mathfrak{S}' \sqcap \mathfrak{S}) \sqcup (\mathfrak{S}' \sqcap \mathfrak{S}^\perp) = \mathbf{0} \sqcup \mathbf{0} = \mathbf{0} \quad . \tag{88}$$

A finite dimensional Hilbert lattice is modular. Since Hilbert lattices are orthomodular lattices, they can be constructed by the pasting of blocks (blocks are maximal Boolean subalgebras); the blocks need not be (almost) disjoint.

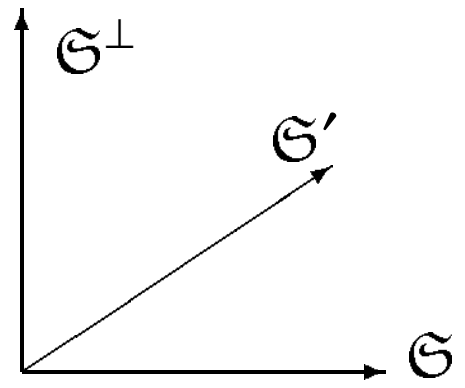

Figure 7: Demonstration of the nondistributivity of Hilbert lattices.

## 2.5 Partial algebras

*Partial algebras* have been introduced by S. Kochen and E. P. Specker [78, 46, 47, 48] as a variant of the classical (Boolean) propositional calculus which takes into account that pairs of propositions may be *incompatible*. A detailed discussion of partial algebras can be found in [17]; connections to quantum logic in [19].

As has been argued before, certain quantum physical statements are no longer simultaneously measurable (cf. compatibility, p. 9). This can be formalized by the introduction of a binary compatibility (commeasurability) relation "$\heartsuit(P_1, P_2)$". Any order relation $a \rightarrow b$ is defined if and only if the propositions $P_1$ and $P_2$ are simultaneously measurable. The propositions $P_1$ and $P_2$ can then be combined by the usual "and" and "or" operations.

More precisely, consider the partial algebra of linear subspaces of $\mathbb{R}^n$ ($n$-dimensional real space) $\mathfrak{B}(\mathbb{R}^n) = \langle B, \heartsuit, \mathbf{0}, \mathbf{1}, \neg, \vee \rangle$, with

***(i)*** $B$ is the set of linear subspaces of $\mathbb{R}^n$;

***(ii)*** $\heartsuit(a, b)$ holds for subspaces $a, b$ if and only if $a$ and $b$ are orthogonal in the sense of elementary geometry, i.e., if there exists a basis of $\mathbb{R}^n$ containing a basis of $a$ and of $b$; (If $a$ is a subspace of $b$, then $\heartsuit(a, b)$ holds.)

***(iii)*** $\mathbf{0}$ is the 0-dimensional, $\mathbf{1}$ is the $n$-dimensional subspace of $\mathbb{R}^n$;

***(iv)*** $\neg a$ is the orthogonal complenet of $a$; and

***(v)*** $a \vee b$ is the linear span of $a$ and $b$, defined only for those pairs $a, b$ for which $\heartsuit(a, b)$ holds.

A well-formed Formula is *valid* if it is valid for all compatibile (commeasurabile) propostions.

Given the concept of partial algebras, it is quite natural to ask whether certain statements which are classical tautologies are still valid in the domain of partial algebras. Furthermore, one may ask whether it is possible to "enrich" the partial algebra of quantum propositions by the introduction of new, "hidden" propositions such that in this enlarged domain the classical (Boolean) algebra is valid. In proving that there exist classical tautologies which are no quantum logical ones, Kochen and Specker gave a negative answer to the latter question [48] (cf. p. 27).

# 3 Quantum information theory

The classical and the quantum mechanical concept of information differ from each other in several aspects. Intuitively and classically, a unit of information is context-free. It is independent of what other information is or might be present. A classical bit remains unchanged, no matter by what methods it is inferred. It can be copied. No doubts can be left.

By contrast, quantum information is contextual. It will be argued below that a quantum bit may appear different, depending on the method by which it is inferred. Quantum bits cannot be copied or "cloned." Classical tautologies are not necessarily satisfied in quantum information theory.

## 3.1 Information is physical

"Information is physical" is the theme of a recent article by Landauer [49], in which lower bounds for the heat dissipation for the processing of classical bits are reviewed. The result can be stated simply by, "there are no unavoidable energy consumption requirements per step in a computer." Only irreversible deletion of classical information is penalized with an increase of entropy.

The slogan "information is physical" is also a often used exclamation in quantum information theory. We not only have to change classical predicate logic in order to make it applicable to (micro-) physics; we have to modify our classical concept of information too.

Classical information theory (e.g., [39]) is based on the bit as fundamental atom. This classical bit, henceforth called *cbit*, is physically represented by one of two classical states of a classical physical system. It is customary to use the symbols "**0**" and "**1**" (interpretable, for instance, as "`false`" and "`true`") as the names of these states. The corresponding classical bit states are $\{\mathbf{0},\mathbf{1}\}$.

In quantum information theory (cf. [1, 4, 24, 34, 62, 5, 56, 25, 26]), the most elementary unit of information, henceforth called *qbit*, may be physically represented by a coherent superposition of the two states which correspond to the symbols **0** and **1**. The qbit states are the coherent superposition of the classical basis states $\{|\,\mathbf{0}\rangle, |\,\mathbf{1}\rangle\}$. They are in the undenumerable set

$$\{|a,b\rangle \mid |a,b\rangle = a|\mathbf{0}\rangle + b|\mathbf{1}\rangle,\ |a|^2+|b|^2=1,\ a,b\in\mathbb{C}\} \quad . \tag{89}$$

## 3.2 Copying and cloning of qbits

Can a qbit be copied? No! — This answer amazes the classical mind.[12] The reason is that any attempt to copy a coherent superposition of states results either in a state reduction, destroying coherence, or, most important of all, in the addition of noise which manifests

[12]Copying of qbits would allow circumvention of the Heisenberg uncertainty relation by measuring two incompatible observables on two identical qbit copies. It would also allow faster-than-light transmission of information [41].

itself as the spontaneous excitations of previously nonexisting field modes [85, 54, 60, 36, 14].

This can be seen by a simple calculation [85]. A physical realization[13] of the qbit state in equation (89) is a two-mode boson field with the identifications

$$|a,b\rangle = a|\mathbf{0}\rangle + b|\mathbf{1}\rangle \quad , \tag{90}$$

$$|\mathbf{1}\rangle = |\,0_1,1_2\rangle \quad , \tag{91}$$

$$|\mathbf{0}\rangle = |\,1_1,0_2\rangle \quad . \tag{92}$$

The classical bit states are $|0_1,1_2\rangle$ and $|1_1,0_2\rangle$. An ideal amplifier, denoted by $|\,A\rangle$, should be able to copy a classical bit state; i.e., it should create an identical particle in the same mode

$$|A_i\rangle|0_1,1_2\rangle \rightarrow |A_f\rangle|0_1,2_2\rangle \quad , \qquad |A_i\rangle|1_1,0_2\rangle \rightarrow |A_f\rangle|2_1,0_2\rangle \quad . \tag{93}$$

Here, $A_i$ and $A_f$ stand for the initial and the final state of the amplifier.

What about copying a true qbit; i.e., a *coherent superposition* of the cbits $|0_1,1_2\rangle$ and $|1_1,0_2\rangle$? According to the quantum evolution law (27), the corresponding amplification process should be representable by a linear (unitary) operator; thus

$$|A_i\rangle(a|0_1,1_2\rangle + b|1_1,0_2\rangle) \rightarrow |A_f\rangle(a|0_1,2_2\rangle + b|2_1,0_2\rangle) \quad . \tag{94}$$

Yet, the true copy of that qbit is the state

$$(a|0_1,1_2\rangle + b|1_1,0_2\rangle)^2 = (a\,a_1^\dagger + b\,a_2^\dagger)^2\,|\,0\rangle = a^2|0_1,2_2\rangle + 2ab|0_1,1_2\rangle|1_1,0_2\rangle + b^2|2_1,0_2\rangle \quad . \tag{95}$$

By comparing (94) with (95) it can be seen that no reasonable (linear unitary quantum mechanical evolution for an) amplifier exists which could copy a generic qbit.

A more detailed analysis (cf. [54, 60], in particular [36, 14]) reveals that the copying (amplification) process generates an amplification of the signal but necessarily adds noise at the same time. This noise can be interpreted as spontaneous emission of field quanta (photons) in the process of amplification.

## 3.3 Context dependence of qbits

Assume that in an EPR-type arragement [30] one wants to measure the product

$$P = m_x^1 m_x^2 m_y^1 m_y^2 m_z^1 m_z^2$$

of the direction of the spin components of each one of the two associated particles 1 and 2 along the $x$, $y$ and $z$-axes. Assume that the operators are normalized such that $|m_i^j| = 1$, $i \in \{x,y,z\}$, $j \in \{1,2\}$. One way to determine $P$ is measuring and, based on these measurements, "counterfactually inferring" [63, 57] the three "observables" $m_x^1 m_y^2$, $m_y^1 m_x^2$ and $m_z^1 m_z^2$. By multiplying them, one obtains +1. Another, alternative, way to determine $P$ is measuring and, based on these measurements, "counterfactually inferring" the three

[13] the most elementary realization is a one-mode field with the symbol **0** corresponding to $|\,0\rangle$ (empty mode) and **1** corresponding to $|\,1\rangle$ (one-quantum filled mode).

"observables" $m^1_x m^2_x$, $m^1_y m^2_y$ and $m^1_z m^2_z$. By multiplying them, one obtains −1. In that way, one has obtained either $P = 1$ or $P = -1$. Associate with $P = 1$ the bit state zero **0** and with $P = -1$ the bit state **1**. Then the bit is either in state zero or one, depending on the way or *context* it was inferred.

This kind of contextuality is deeply rooted in the non-Boolean algebraic structure of quantum propositions. Note also that the above argument relies heavily on counterfactual reasoning, because, for instance, only two of the six observables $m^j_i$ can actually be experimentally determined.

## 3.4 Classical versus quantum tautologies

I shall review the shortest example of a classical tautology which is not valid in threedimensional (real) Hilbert space that is known up-to-date [71].

Consider the propositions

$$
\begin{aligned}
d_1 &\rightarrow \neg b_2 \quad , & (96)\\
d_1 &\rightarrow \neg b_3 \quad , & (97)\\
d_2 &\rightarrow a_2 \vee b_2 \quad , & (98)\\
d_2 &\rightarrow \neg b_3 \quad , & (99)\\
d_3 &\rightarrow \neg b_2 \quad , & (100)\\
d_3 &\rightarrow (a_1 \vee a_2 \rightarrow b_3) \quad , & (101)\\
d_4 &\rightarrow a_2 \vee b_2 \quad , & (102)\\
d_4 &\rightarrow (a_1 \vee a_2 \rightarrow b_3) \quad , & (103)\\
(a_2 &\vee c_1) \vee (b_3 \vee d_1) \quad , & (104)\\
(a_2 &\vee c_2) \vee (a_1 \vee b_1 \rightarrow d_1) \quad , & (105)\\
c_1 &\rightarrow b_1 \vee d_2 \quad , & (106)\\
c_2 &\rightarrow b_3 \vee d_2 \quad , & (107)\\
(a_2 &\vee c_1) \vee [(a_1 \vee a_2 \rightarrow b_3) \rightarrow d_3] \quad , & (108)\\
(a_2 &\vee c_2) \vee (b_1 \vee d_3) \quad , & (109)\\
c_2 &\rightarrow [(a_1 \vee a_2 \rightarrow b_3) \rightarrow d_4] \quad , & (110)\\
c_1 &\rightarrow (a_1 \vee b_1 \rightarrow d_4) \quad , & (111)\\
(a_1 &\rightarrow a_2) \vee b_1 \quad . & (112)
\end{aligned}
$$

The proposition formed by $F$ : (96)$\wedge\cdots\wedge$(111)$\rightarrow$(112) is a classical tautology.

$F$ is not valid in threedimensional (real) Hilbert space $E^3$, provided one identifies the $a$'s, $b$'s and $c$'s with the following onedimensional subspaces of $E^3$:

$$
\begin{aligned}
a_1 &\equiv \mathfrak{S}(1,0,0) \quad , & (113)\\
a_2 &\equiv \mathfrak{S}(0,1,0) \quad , & (114)\\
b_1 &\equiv \mathfrak{S}(0,1,1) \quad , & (115)\\
b_2 &\equiv \mathfrak{S}(1,0,1) \quad , & (116)
\end{aligned}
$$

$$b_3 \equiv \mathfrak{S}(1,1,0) \quad , \tag{117}$$

$$c_1 \equiv \mathfrak{S}(1,0,2) \quad , \tag{118}$$

$$c_2 \equiv \mathfrak{S}(2,0,1) \quad , \tag{119}$$

$$d_1 \equiv \mathfrak{S}(-1,1,1) \quad , \tag{120}$$

$$d_2 \equiv \mathfrak{S}(1,-1,1) \quad , \tag{121}$$

$$d_3 \equiv \mathfrak{S}(1,1,-1) \quad , \tag{122}$$

$$d_4 \equiv \mathfrak{S}(1,1,1) \quad , \tag{123}$$

where $\mathfrak{S}(v) = \{av \mid a \in \mathbb{R}\}$ is the subspace spanned by $v$.

Let the "or" operation be represented by $\mathfrak{S}(v) \vee \mathfrak{S}(w) = \{av + bw \mid a, b \in \mathbb{R}\}$ the linear span of $\mathfrak{S}(v)$ and $\mathfrak{S}(w)$.

Let the "and" operation be represented by $\mathfrak{S}(v) \wedge \mathfrak{S}(w) = \mathfrak{S}(v) \cap \mathfrak{S}(w)$ the set theoretic complement of $\mathfrak{S}(v)$ and $\mathfrak{S}(w)$.

Let the complement be represented by $\neg\mathfrak{S}(v) = \{w \mid v \cdot w = 0\}$ the orthogonal subspace of $\mathfrak{S}(v)$.

Let the "implication" relation be represented by $\mathfrak{S}(v) \rightarrow \mathfrak{S}(w) \equiv (\neg\mathfrak{S}(v)) \vee \mathfrak{S}(w)$.

Then, (96), $\cdots$, (111)= $E^3$, whereas (112)= $\neg\mathfrak{S}(1,0,0) \neq E^3$. Therefore, at least for states lying in the direction $(1,0,0)$ [18], $F$ is not a quantum tautology.

The set of eleven rays can be represented by vectors from the center of a cube to the indicated points [63], as drawn in Fig. 8.

# 4 Elements of quantum computatability and complexity theory

Can a quantum computer do what a classical one cannot do?

Notice that the concept of universal computation is based on *primary intuitive concepts* of "reasonable" instances of "mechanic" computation, which refer to *physical insight;* i.e., which refer to the types of processes which can be performed in the physical world. In this sense, the level of physical comprehension sets the limits to whatever is acceptable as valid method of computation. If, for instance, we could employ an "oracle" (the term "oracle" denotes a subprogram which supplies the true answer to a problem, if a true answer "platonically" exists), our wider notion of "mechanic" computation would include oracle computation. Likewise, a computationally more restricted universe would intrinsically imply a more restricted concept of "mechanic" computation. For an early and brilliant discussion of this aspect the reader is referred to A. M. Turing's original work [82]. As D. Deutsch puts it ([24], p. 101),

> *"The reason why we find it possible to construct, say, electronic calculators, and indeed why we can perform mental arithmetic, cannot be found in mathematics or logic.* The reason is that the laws of physics 'happen to' permit the existence of physical models for the operations of arithmetic *such as addition, subtraction and multiplication. If they did not, these familiar*

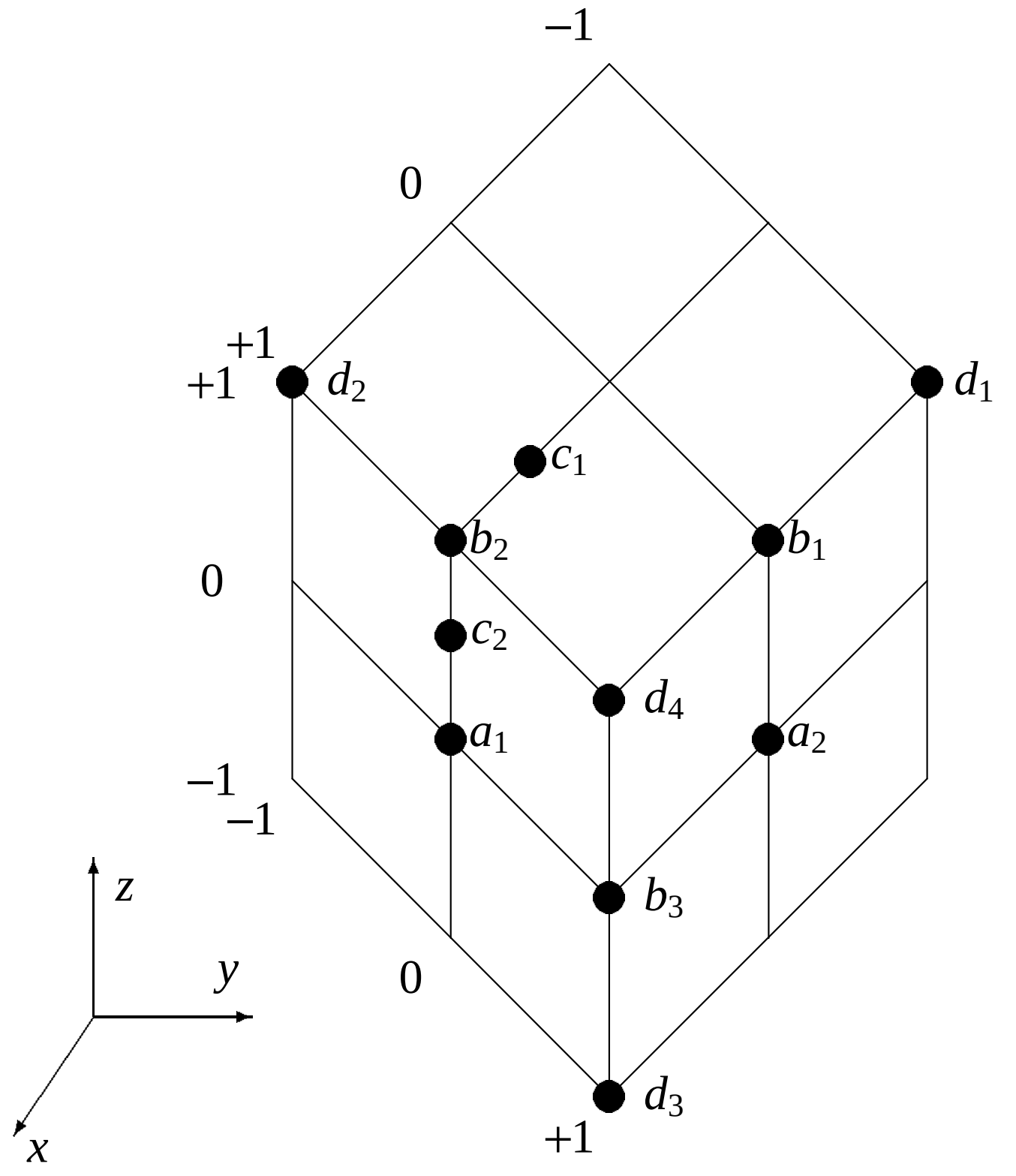

Figure 8: The eleven rays in the proof of the Kochen-Specker theorem based on the construction of Schütte are obtained by connecting the center of the cube to the black dots on its faces and edges.

*operations would be non-computable functions. We might still know* of *them and invoke them in mathematical proofs (which would presumably be called 'non constructive') but we could not perform them."*

For another discussion of this topic, see R. Rosen [69] and M. Davis' book [21], p. 11, where the following question is asked:

*" … how can we ever exclude the possibility of our presented, some day (perhaps by some extraterrestrial visitors), with a (perhaps extremely complex) device or "oracle" that "computes" a non computable function?"*

Indeed, the concept of *Turing degree* [82] yields an extension of universal computation to oracle computation. In a way, the syntactic aspect of computation could thus be "tuned" to wider concepts of (physically realisable) computation if that turns out to be necessary. As has been pointed out before, such a change cannot be decided by syntactic reasoning; it has to be motivated by physical arguments. Thus, at least at the level of investigation of "reasonable" instances of computation, the theory of computable functions, recursion theory, is part of physics.

The following features are necessary but not sufficient qualities of quantum computers.

- Input, output, program and memory are qbits; the basis $\{ \mid \mathbf{0}\rangle, \mid \mathbf{1}\rangle\}$ represent the classical bit values **0** and **1**;
- any computation (step) can be represented by a unitary transformation of the computer as a whole;
- quantum measurements of the basis states $\mid \mathbf{0}\rangle$ and $\mid \mathbf{1}\rangle$ may be carried out on any qbit at any stage of the computation;
- because of the unitarity of the quantum evolution operator, any computation is reversible; Therefore, a deterministic computation can be performed by a quantum computer if and only if it is reversible, i.e., if the program does not involve "deletion" or "erasure" of information (cf. [49]);
- at the end of the computation a measurement is made; The measurement is usually irreversible, and the computer is in a classical bit state.

## 4.1 Universal quantum computers

In what follows, a Turing machine will be quantized. This means that all entities of the Turing machine will be described in quantum mechanical, i.e., Hilbert space, terminology.

Assume a universal computer $\mathfrak{U}$ consistsing of a finite processor and an infinite (tape) memory [4, 24]. Its state function $\mid \psi_{\mathfrak{U}}\rangle$ can be represented as follows. Assume further that the processor consists of $M$ 2-state observables

$$\hat{p}_i \quad , \quad i \in \mathbb{Z}_M = \{0, 1, 2, 3, \ldots, M-1\} \quad . \tag{124}$$

The memory consists of an infinite sequence of 2-state observables

$$\hat{m}_i \quad , \quad i \in \mathbb{Z} = \{0,1,2,3,\ldots\} \quad . \tag{125}$$

These infinite sequence may be thought of as modelling $\mathfrak{U}$'s infinite tape. Corresponding to the tape position another operator

$$\hat{x} = x \in \mathbb{Z} = \{0,1,2,3,\ldots\} \quad . \tag{126}$$

is defined. Let $x, \vec{n}, \vec{m}$ be the eigenvalues of $\hat{x}, \hat{\vec{n}}, \hat{\vec{m}}$, respectively. The Turing machine's state function $\mid \psi_{\mathfrak{U}}\rangle$ can then be written as a unit vector in the infinite-dimensional Hilbert space $\mathfrak{H}_{\mathfrak{U}}$ spanned by the simultaneous eigenvectors

$$\mid \psi_{\mathfrak{U}}\rangle \equiv \mid x\rangle \otimes \mid \vec{n}\rangle \otimes \mid \vec{m}\rangle = \mid x;\vec{n};\vec{m}\rangle = \mid x; n_1, n_2, n_3, \ldots, n_M; m_1, m_2, m_3, \ldots\rangle \quad . \tag{127}$$

The states $\mid x;\vec{n};\vec{m}\rangle$ defined in (127) are the *computational basis states.*

The process of computation can be described as follows. Assume that initially, i.e., at time $t = 0$, the computer is at position $x = 0$ and has a "blank" processor $\vec{n} = \vec{0}$. The tape's quantum numbers $\vec{m}$ characterize the "program" and the "input" of $\mathfrak{U}$. A quantum computer can be in a coherent superposition of such states

$$\mid \psi_{\mathfrak{U}}(t=0)\rangle = \sum_{\vec{m}} a_{\vec{m}} \mid 0, \vec{0}, \vec{m}\rangle \quad , \quad \sum_{\vec{m}} |a_{\vec{m}}|^2 = 1 \quad . \tag{128}$$

Let $\hat{U}$ be the linear unitary evolution operator corresponding to $\mathfrak{U}$. The dynamics is discrete; i.e., in time steps $T$

$$\mid \psi_{\mathfrak{U}}(nT)\rangle = \hat{U}^n \mid \psi_{\mathfrak{U}}(0)\rangle \quad , \quad n \in \mathbb{Z}^n = \{0,1,2,3,\ldots\} \quad . \tag{129}$$

Tunring machines operate "locally;" i.e., their tape memory can only shift one position at a single computation step. Furthermore, only the $x$'th bit of the memory tape participates in a single computation step. This can be implemented by

$$\begin{aligned}\langle x;\vec{n}';\vec{m}' \mid U \mid x;\vec{n};\vec{m}\rangle \quad = \quad & \{\delta_{x+1\, x'} U^+(|n', m'_x; |n, m_x) + \\ & \delta_{x\, x'} U^0(|n', m'_x; |n, m_x) + \\ & \delta_{x-1\, x'} U^-(|n', m'_x; |n, m_x)\} \prod_{y \neq x} \delta_{m_y\, m'_y} \quad .\end{aligned} \tag{130}$$

## 4.2 Universal quantum networks

The "brute force" method of obtaining a (universal) quantum computer by quantizing the "hardware" components of a Turing machine seems to suffer from the same problem as its classical counterpart. — It seems technologically unreasonable to actually construct a universal quantum device with a "scaled down" (to nanometer size) model of a Turing machine in mind.[14]

[14] I would suspect that future historians of science will construct such a device; just as Babbage's calculater has been constructed by staff of the British Museum recently.

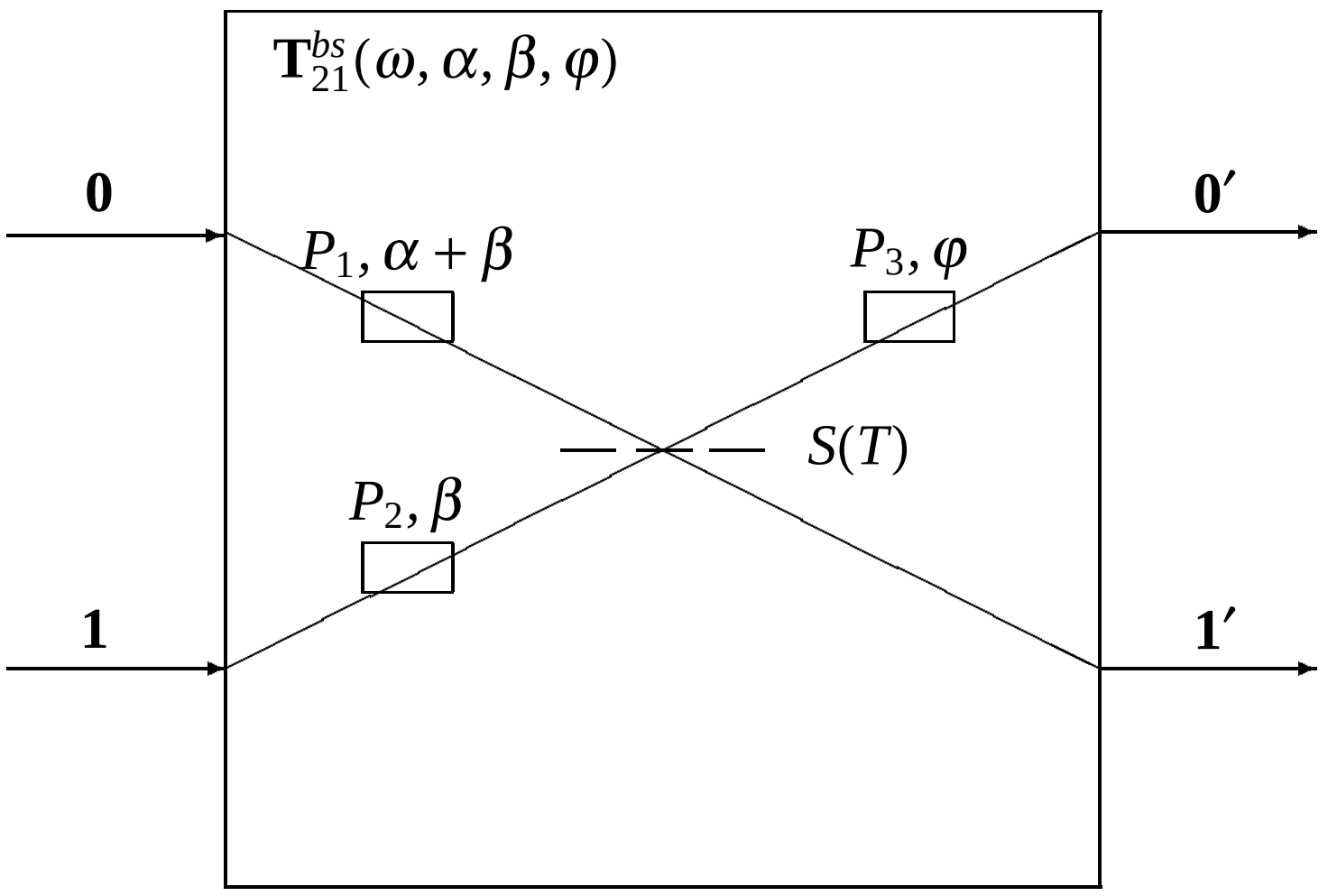

a)

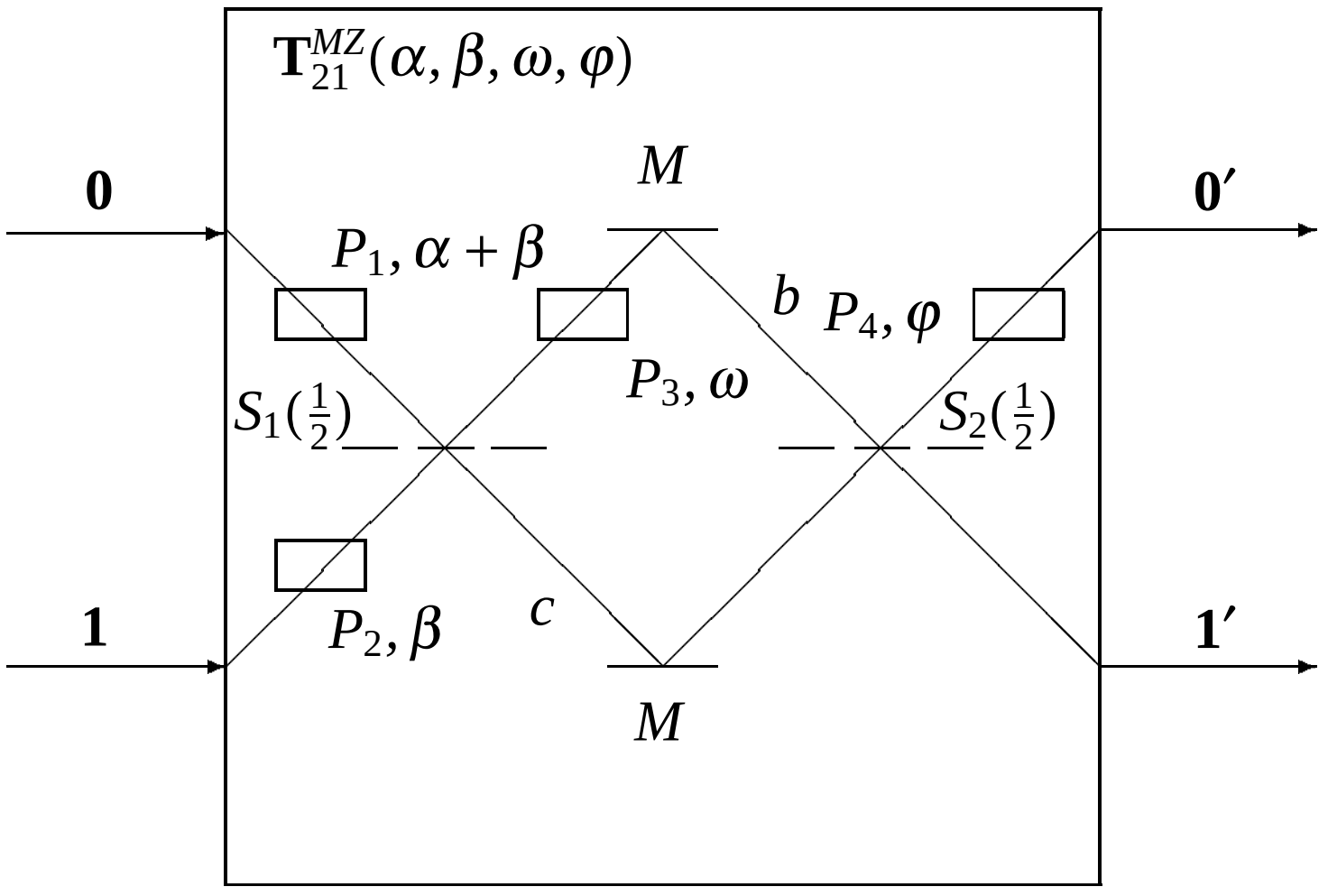

b)

Figure 9: Elementary quantum interference device. An elementary quantum interference device can be realized by a 4-port interferometer with two input ports $\mathbf{0}, \mathbf{1}$ and two output ports $\mathbf{0}', \mathbf{1}'$. Any twodimensional unitary transformation can be realised by the devices. a) shows a realization by a single beam splitter $S(T)$ with variable transmission $t$ and three phase shifters $P_1, P_2, P_3$; b) shows a realization with 50:50 beam splitters $S_1(\frac{1}{2})$ and $S_2(\frac{1}{2})$ and four phase shifters $P_1, P_2, P_3, P_4$.

One of the novel physical features of quantum information theory is the superposition of classical information. We shall therefore consider quantum interference devices such as the beam splitter or the Mach-Zehnder interferometer (cf. 18).

The elementary quantum interference device $\mathbf{T}_{21}^{ps}$ depicted in Fig. (9.a) is just a beam splitter followed by a phase shifter in one of the output ports. According to the "toolbox" rules (cf. p. 19), the process can be quantum mechanically described by[15]

$$
\begin{aligned}
P_1: \; |\mathbf{0}\rangle &\rightarrow |\mathbf{0}\rangle e^{i\alpha+\beta} \quad , && (131)\\
P_2: \; |\mathbf{1}\rangle &\rightarrow |\mathbf{1}\rangle e^{i\beta} \quad , && (132)\\
S: \; |\mathbf{0}\rangle &\rightarrow T\,|\mathbf{1}'\rangle + iR\,|\mathbf{0}'\rangle \quad , && (133)\\
S: \; |\mathbf{1}\rangle &\rightarrow T\,|\mathbf{0}'\rangle + iR\,|\mathbf{1}'\rangle \quad , && (134)\\
P_3: \; |\mathbf{0}'\rangle &\rightarrow |\mathbf{0}'\rangle e^{i\varphi} \quad . && (135)
\end{aligned}
$$

If $|\,\mathbf{0}\rangle \equiv |\,\mathbf{0}'\rangle \equiv \begin{pmatrix} 1 \\ 0 \end{pmatrix}$ and $|\,\mathbf{1}\rangle \equiv |\,\mathbf{1}'\rangle \equiv \begin{pmatrix} 0 \\ 1 \end{pmatrix}$ and $R(\omega) = \sin\omega$, $T(\omega) = \cos\omega$, then the corresponding unitary evolution matrix which transforms any coherent superposition of $|\,\mathbf{0}\rangle$ and $|\,\mathbf{1}\rangle$ into a superposition of $|\,\mathbf{0}'\rangle$ and $|\,\mathbf{1}'\rangle$ is given by

$$
\begin{aligned}
\mathbf{T}_{21}^{bs}(\omega,\alpha,\beta,\varphi) &= \left[ e^{i\beta} \begin{pmatrix} i\,e^{i(\alpha+\varphi)}\,\sin\omega & e^{i\alpha}\,\cos\omega \\ e^{i\varphi}\,\cos\omega & i\,\sin\omega \end{pmatrix} \right]^{-1} \\
&= e^{-i\beta} \begin{pmatrix} -i\,e^{-i(\alpha+\varphi)}\,\sin\omega & e^{-i\varphi}\,\cos\omega \\ e^{-i\alpha}\,\cos\omega & -i\,\sin\omega \end{pmatrix} \quad . \qquad (136)
\end{aligned}
$$

The elementary quantum interference device $\mathbf{T}_{21}^{MZ}$ depicted in Fig. (9.b) is a (rotated) Mach-Zehnder interferometer with *two* input and output ports and three phase shifters. According to the "toolbox" rules, the process can be quantum mechanically described by

$$
\begin{aligned}
P_1: \; |\mathbf{0}\rangle &\rightarrow |\mathbf{0}\rangle e^{i\alpha+\beta} \quad , && (137)\\
P_2: \; |\mathbf{1}\rangle &\rightarrow |\mathbf{1}\rangle e^{i\beta} \quad , && (138)\\
S_1: \; |\mathbf{1}\rangle &\rightarrow (|b\rangle + i\,|c\rangle)/\sqrt{2} \quad , && (139)\\
S_1: \; |\mathbf{0}\rangle &\rightarrow (|c\rangle + i\,|b\rangle)/\sqrt{2} \quad , && (140)\\
P_3: \; |c\rangle &\rightarrow |c\rangle e^{i\omega} \quad , && (141)\\
S_2: \; |b\rangle &\rightarrow (|\mathbf{1}'\rangle + i\,|\mathbf{0}'\rangle)/\sqrt{2} \quad , && (142)\\
S_2: \; |c\rangle &\rightarrow (|\mathbf{0}'\rangle + i\,|\mathbf{1}'\rangle)/\sqrt{2} \quad , && (143)\\
P_4: \; |\mathbf{0}'\rangle &\rightarrow |\mathbf{0}'\rangle e^{i\varphi} \quad . && (144)
\end{aligned}
$$

When again $|\,\mathbf{0}\rangle \equiv |\,\mathbf{0}'\rangle \equiv \begin{pmatrix} 1 \\ 0 \end{pmatrix}$ and $|\,\mathbf{1}\rangle \equiv |\,\mathbf{1}'\rangle \equiv \begin{pmatrix} 0 \\ 1 \end{pmatrix}$, then the corresponding unitary evolution matrix which transforms any coherent superposition of $|\,\mathbf{0}\rangle$ and $|\,\mathbf{1}\rangle$ into a

[15] Alternatively, the action of a lossless beam splitter may be described by the matrix $\begin{pmatrix} T(\omega) & i\,R(\omega) \\ i\,R(\omega) & T(\omega) \end{pmatrix} = \begin{pmatrix} \cos\omega & i\,\sin\omega \\ i\,\sin\omega & \cos\omega \end{pmatrix}$. A phase shifter in twodimensional Hilberts space is represented by either $\begin{pmatrix} e^{i\varphi} & 0 \\ 0 & 1 \end{pmatrix}$ or $\begin{pmatrix} 1 & 0 \\ 0 & e^{i\varphi} \end{pmatrix}$. The action of the entire device consisting of such elements is calculated by multiplying the matrices in reverse order in which the quanta pass these elements [86, 13].

superposition of $|\,\mathbf{0}'\rangle$ and $|\,\mathbf{1}'\rangle$ is given by

$$\mathbf{T}_{21}^{MZ}(\alpha,\beta,\omega,\varphi) = -i\,e^{-i(\beta+\frac{\omega}{2})}\begin{pmatrix} -e^{-i\,\alpha+\varphi}\sin(\frac{\omega}{2}) & e^{-i\,\varphi}\cos(\frac{\omega}{2}) \\ e^{-i\,\alpha}\cos(\frac{\omega}{2}) & \sin(\frac{\omega}{2}) \end{pmatrix} \quad . \tag{145}$$

The correspondence between $\mathbf{T}_{21}^{bs}(T(\omega),\alpha,\beta,\varphi)$ with $\mathbf{T}_{21}^{MZ}(\alpha',\beta',\omega',\varphi')$ in equations (136) (145) can be verified by comparing the elements of these matrices. The resulting four equations can be used to eliminate the four unknown parameters $\omega' = 2\omega$, $\beta' = \beta-\omega$, $\alpha' = \alpha - \pi/2$, $\beta' = \beta - \omega$ and $\varphi' = \varphi - \pi/2$; i.e.,

$$\mathbf{T}_{21}^{bs}(T(\omega),\alpha,\beta,\varphi) = \mathbf{T}_{21}^{MZ}(\alpha - \frac{\pi}{2},\beta-\omega,2\omega,\varphi-\frac{\pi}{2}) \quad . \tag{146}$$

Both elementary quantum interference devices are *universal* in the sense that *every* unitary quantum evolution operator in twodimensional Hilbert space can be brought into a one-to-one correspondence to $\mathbf{T}_{21}^{bs}$ and $\mathbf{T}_{21}^{MZ}$; with corresponding values of $T,\alpha,\beta,\varphi$ or $\alpha,\omega,\beta,\varphi$. This can be easily seen by a similar calculation as before; i.e., by comparing equations (136) (145) with the "canonical" form of a unitary matrix, which is the product of a $U(1) = e^{-i\,\beta}$ and of the unimodular unitary matrix $SU(2)$ [59]

$$\mathbf{T}(\omega,\alpha,\varphi) = \begin{pmatrix} e^{i\,\alpha}\cos(\omega) & -e^{-i\,\varphi}\sin(\omega) \\ e^{i\,\varphi}\sin(\omega) & e^{-i\,\alpha}\cos(\omega) \end{pmatrix} \quad , \tag{147}$$

where $-\pi \le \beta,\omega \le \pi$, $-\frac{\pi}{2}\alpha,\varphi \le \frac{\pi}{2}$. Let

$$\mathbf{T}(\omega,\alpha,\beta,\varphi) = e^{-i\,\beta}\mathbf{T}(\omega,\alpha,\varphi) \quad . \tag{148}$$

A proper identification of the parameters $\alpha,\beta,\omega,\varphi$ yields

$$\mathbf{T}(\omega,\alpha,\beta,\varphi) = \mathbf{T}_{21}^{bs}(\omega - \frac{\pi}{2},-\alpha-\varphi-\frac{\pi}{2},\beta+\alpha+\frac{\pi}{2},\varphi-\alpha+\frac{\pi}{2}) \quad . \tag{149}$$

Let us examine the realization of a few primitive logical "gates" corresponding to unary operations on qbits. The "identity" element $\mathbb{I}$ is defined by $|\,\mathbf{0}\rangle \to |\,\mathbf{0}\rangle$, $|\,\mathbf{1}\rangle \to |\,\mathbf{1}\rangle$ and can be realized by

$$\mathbb{I} = T_{21}^{bs}(-\frac{\pi}{2},-\frac{\pi}{2},\frac{\pi}{2},\frac{\pi}{2}) = T_{21}^{MZ}(-\pi,\pi,-\pi,0) = \begin{pmatrix} 1 & 0 \\ 0 & 1 \end{pmatrix} \quad . \tag{150}$$

The "$\texttt{not}$" element is defined by $|\,\mathbf{0}\rangle \to |\,\mathbf{1}\rangle$, $|\,\mathbf{1}\rangle \to |\,\mathbf{0}\rangle$ and can be realized by

$$\texttt{not} = T_{21}^{bs}(0,0,0,0) = T_{21}^{MZ}(-\frac{\pi}{2},0,0,-\frac{\pi}{2}) = \begin{pmatrix} 0 & 1 \\ 1 & 0 \end{pmatrix} \quad . \tag{151}$$

The next element, "$\sqrt{\texttt{not}}$" is a truly quantum mechanical; i.e., nonclassical, one, since it converts a classical bit into a coherent superposition of $|\,\mathbf{0}\rangle$ and $|\,\mathbf{1}\rangle$. $\sqrt{\texttt{not}}$ is defined by $|\,\mathbf{0}\rangle \to |\,\mathbf{0}\rangle + |\,\mathbf{1}\rangle$, $|\,\mathbf{1}\rangle \to -|\,\mathbf{0}\rangle + |\,\mathbf{1}\rangle$ and can be realized by

$$\sqrt{\texttt{not}} = T_{21}^{bs}(-\frac{\pi}{4},-\frac{\pi}{2},\frac{\pi}{2},\frac{\pi}{2}) = T_{21}^{MZ}(-\pi,\frac{3\pi}{4},-\frac{\pi}{2},0) = \frac{1}{\sqrt{2}}\begin{pmatrix} 1 & -1 \\ 1 & 1 \end{pmatrix} \quad . \tag{152}$$

Note that $\sqrt{\texttt{not}} \cdot \sqrt{\texttt{not}} = \texttt{not}$.

It is very important that the elementary quantum interference device realizes an arbitrary quantum time evolution of a twodimensional system. The performance of the quantum interference device is determined by four parameters, corresponding to the phases $\alpha, \beta, \varphi, \omega$.

Any $n$-dimensional unitary matrix $U$ can be composed from elementary unitary transformations in twodimensional subspaces of $\mathbb{C}^n$. This is usually shown in the context of parametrization of the $n$-dimensional unitary groups (cf. [59], chapter 2 and [67]). The number elementary transformations is polynomially bounded and does not exceed $\binom{n}{2} = \frac{n(n-1)}{2} = O(n^2)$.

Thus a suitable arrangement of elementary quantum interference devices — representing unitary transformations on twodimensional subspaces of the computational basis states [cf. equation (127)] — forms a universal quantum network [34, 24].

## 4.3 Quantum recursion theory

This section deals mainly with Cantor's method of diagonalization, which is applied for undecidability proofs in recursion theory [68, 61]. Due to the possibility of a coherent superposition of classical bit states, the usual *reductio ad absurdum* argument breaks down. Instead, diagonalization procedures in quantum information theory yield qbit solutions which are fixed points of the associated unitary operators.

I shall demonstrate the emergence of fixed points by a simple example. Diagonalization effectively transforms the classical bit value "**0**" into value "**1**" and "**1**" into "**0**." The evolution representing diagonaliation can be expressed by the unitary operator $\widehat{D}$ as follows $\widehat{D}|\mathbf{0}\rangle \rightarrow |\mathbf{1}\rangle$, and $\widehat{D}|\mathbf{1}\rangle \rightarrow |\mathbf{0}\rangle$. In the state basis $|\,\mathbf{0}\rangle \equiv \begin{pmatrix} 1 \\ 0 \end{pmatrix}$ and $|\,\mathbf{1}\rangle \equiv \begin{pmatrix} 0 \\ 1 \end{pmatrix}$ ($\tau_1$ stands for the Pauli spin operator),

$$\widehat{D} = \tau_1 = \texttt{not} = \begin{pmatrix} 0 & 1 \\ 1 & 0 \end{pmatrix} = |\mathbf{1}\rangle\langle\mathbf{0}| + |\mathbf{0}\rangle\langle\mathbf{1}| \quad . \tag{153}$$

$\widehat{D}$ will be called *diagonalization* operator, despite the fact that the only nonvanishing components are off-diagonal.

As has been pointed out earlier, quantum information theory allows a coherent superposition $|\,a, b\rangle = a\ |\,\mathbf{0}\rangle + b\ |\,\mathbf{1}\rangle$ of the classical bit states. Therefore, $\widehat{D}$ has a fixed point at

$$|\,\frac{1}{\sqrt{2}}, \frac{1}{\sqrt{2}}\rangle = \frac{1}{\sqrt{2}}|\mathbf{0}\rangle + \frac{1}{\sqrt{2}}|\mathbf{1}\rangle \quad , \tag{154}$$

which is a coherent superposition of the classical bit base and does not give rise to inconsistencies [81].

Classical undecidability is recovered if one performs any irreversible measurement on the fixed point state. This causes a state reduction into the classical states $|\,\mathbf{0}\rangle$ and $|\,\mathbf{1}\rangle$. The probability for the fixed point state $|\frac{1}{\sqrt{2}}, \frac{1}{\sqrt{2}}\rangle$ to be in either $|\mathbf{0}\rangle$ or $|\mathbf{1}\rangle$ is equal; i.e.,

$$|\langle\mathbf{0}\,|\,\frac{1}{\sqrt{2}}, \frac{1}{\sqrt{2}}\rangle|^2 = |\langle\mathbf{1}\,|\,\frac{1}{\sqrt{2}}, \frac{1}{\sqrt{2}}\rangle|^2 = \frac{1}{2} \quad . \tag{155}$$

However, no contradiction is involved in the diagonalization argument. Therefore, standard proofs of the recursive unsolvability of the halting problem do not apply to quantum recursion theory.

Another, less abstract, application for quantum information theory is the handling of inconsistent information in databases. Thereby, two contradicting cbits of information $|a\rangle$ and $|b\rangle$ are resolved by the qbit $|\frac{1}{\sqrt{2}}, \frac{1}{\sqrt{2}}\rangle = \frac{1}{\sqrt{2}}(|a\rangle + |b\rangle)$. Throughout the rest of the computation the coherence is maintained. After the processing, the result is obtained by an irreversible measurement. The processing of qbits requires an exponential space overhead on classical computers in cbit base [33]. Thus, in order to remain tractable, the corresponding qbits should be implemented on truly quantum universal computers.

## 4.4 Factoring

One decisive features of quantum computation is *parallelism*. During a cycle of computation, a quantum computer proceeds down all coherent paths at once. It should be mentioned that quantum parallelism is a quite different feature from quantum stochasticity. Quantum mechanics has an ideal "build in" randum number generator [80] which can be used by stochastic algorithms. The crucial question is: what can we do with quantum parallelism what we cannot do with classical devices [26, 27, 8, 9, 6, 16, 76]?

A recent proposal by Shor [76] seems to indicate that the high expectations raised by quantum computing are justified. Shor's claim is that quantum factoring and discrete logarithms can be performed in a time which is polynomially (in the number of digits of the input number) bounded on quantum computers. At the heart of Shor's algorithm is a Fourier transformation [20]

$$| a\rangle = \frac{1}{\sqrt{q}} \sum_{b=0}^{q-1} | b\rangle e^{2\pi i ab/q} \quad , \tag{156}$$

where $0 \le a < q$ (the number of bits in $q$ is polynomial). Equation (156) defines a unitary transformation $\langle a \mid A_q \mid b\rangle = (1/\sqrt{q}) \exp(2\pi i ab/q)$. $\langle a \mid A_q \mid b\rangle$ can be computed in polynomial time on a quantum computer. The reason for this is that its elements are evaluated "all at once" in parallel, a genuine quantum feature.

The unitary operator $\hat{A}_g$ is used as an (polynomial-time) "oracle" for the computation of the order of an element $x$ in the multiplicative group (mod $n$); i.e., the least integer $r$ such that $x^r \equiv 1 (\text{mod } x)$. There is a randomized reduction from factoring to the order of an element.

## 4.5 Travelling salesman

Unlike factoring, which seems to be situated "inbetween" the classical polynomial-time and *NP*-complete problems (presently, the best classical estimate for the computing time for factoring is $n^{\log\log n}$, where $n$ is the number of digits), the travelling salesman problem is *NP*-complete.

Almost unnoticed by the quantum computing community, Černý has put forward a scheme for the solution of the travelling salesman problem [16]. Assume that the number of cities is $n$. Černý's quantum computer consists of a series of $n-1$ slit layers (for cities number $2, 3, \ldots, n$) for quantum interference. Each one of the layers has $n-1$ slits (for cities number $2, 3, \ldots, n$). The array of slits is irradiated by a beam of quanta (e.g., light, neutrons, electrons). There are $(n-1)^{n-1}$ possible classical trajectories for any single classical particle. This means that in order to exhaust all trajectories, one needs $(n-1)^{n-1}$ classical paricles. But quanta are no classical entities. By the coherent superposition of trajectories, every quantum has a nonvanishing amplitude to pass *all* the $(n-1)^{n-1}$ trajectories in polynomial time $O(n)$. Stated pointedly, the quantum "experiences" exponentially many paths in polynomial time. One may therefore hope to be able to detect the shortest path by a "suitable" measurement of the corresponding observable of the quantum.

Now, what can we make from that? At first glance it surely sounds like a "cheap lunch!" But Černý clearly states that the price is high indeed as far as statistics is concerned: in order to obtain results with reasonable probabilities, one has to dedicate of the order of $N!$ quanta to the task, resulting in a non-polynomial increase in energy. (This might be exactly the case for classical analog devices.)

## 4.6 Will the strong Church-Turing thesis survive?

The strong Church-Turing thesis postulates that the class of polynomial-time algorithms is robust; i.e., invariant with respect to variations of "reasonable" models of computation. Quantum complexity theory challenges this claim.[16]

At the heart of the speedup is quantum parallelism. Roughly stated, quantum parallelism assures that a single quantum bit, henceforth called *qbit*, can "branch off" into an arbitrary number of coherent entangled qbits. A typical physical realization of a qbit is a single field mode of a photon (electron, neutron), with the empty and the one-photon state $|\,\mathbf{0}\rangle$ and $|\,\mathbf{1}\rangle$ representing the classical symbols **0** and **1**, respectively. The branching process into coherent beam paths can be realized by an array of beam splitters such as semitransparent mirrors or a double slit. A typical cascade of branching process into $n^k$ coherent beam paths is described by a successive array of $k$ identical beam splitters with $n$ slots and vanishing relative phases

$$|\,s_0\rangle \quad\rightarrow\quad \frac{1}{\sqrt{n}}\left(|\,s_0 s_{11}\rangle + |\,s_0 s_{12}\rangle + \cdots + |\,s_0 s_{1n}\rangle\right) \quad , \tag{157}$$

$$\frac{1}{\sqrt{n}}|\,s_0 s_{11}\rangle \quad\rightarrow\quad \frac{1}{n}\left(|\,s_0 s_{11} s_{21}\rangle + |\,s_0 s_{11} s_{22}\rangle + \cdots + |\,s_0 s_{11} s_{2n}\rangle\right) \tag{158}$$

$$\frac{1}{\sqrt{n}}|\,s_0 s_{12}\rangle \quad\rightarrow\quad \frac{1}{n}\left(|\,s_0 s_{12} s_{21}\rangle + |\,s_0 s_{12} s_{22}\rangle + \cdots + |\,s_0 s_{12} s_{2n}\rangle\right) \tag{159}$$

$$\vdots$$

$$\frac{1}{n^{-(k-1)/2}}|\,s_0 s_{1n}\cdots s_{(k-1)n}\rangle \quad\rightarrow\quad \frac{1}{n^{-k/2}}\left(|\,s_0 s_{1n}\cdots s_{kn}\rangle + \cdots + |\,s_0 s_{1n}\cdots s_{kn}\rangle\right) \ . \tag{160}$$

Notice that every beam splitter contributes a normalization factor of $1/\sqrt{n}$ to the amplitude of the process. The probability amplitude for a single quantum in state $|\,s_0\rangle$ to evolve into

[16]Quantum complexity theory, however, does *not* challenge the Church-Turing thesis; i.e., all quantum-computable objects are classically computable.

one particular beam path $s_0 s_{1i_1} s_{2i_2} s_{3i_3} \cdots s_{ki_k}$ therefore is

$$\langle s_0 s_{1i_1} s_{2i_2} s_{3i_3} \cdots s_{ki_k} \mid U \mid s_0 \rangle = n^{-k/2} \quad , \tag{161}$$

where $U$ stands for the unitary evolution operator corresponding to the array of beam splitters.

More generally, any one of the entangled qbits originating from the branching process can be processed in parallel. The beam path $s_0 s_{1i_1} s_{2i_2} s_{3i_3} \cdots s_{ki_k}$ can be interpreted as a *program code* [39, 15, 12, 80]. How many programs can be coded into one beam path? Notice that, in order to maintain coherence, no code of a valid program can be the prefix of a code of another valid program. Therefeore, in order to maintain the parallel quality of quantum computation, only *prefix* or *instantaneous* codes are allowed. A straightforward proof using induction [39] shows that the instantaneous code of $q$ programs $\{p_1, p_2, \ldots, p_q\}$ with length $l_1 \leq l_2 \leq \cdots \leq l_q$ satisfies the *Kraft inequality*

$$\sum_{i=1}^{q} n^{-l_i} \leq 1 \quad , \tag{162}$$

where $n$ is the number of symbols of the code alphabet. In our case, $n$ is identified with the number of slits in the beam splitters. Stated pointedly, instantaneous decodability restricts the number of legal programs due to the condition that to legal program can be the prefix of another legal program. The Kraft inequality then states that no more than maximally $q = n^k$ programs can be coded by a successive array of $k$ identical beam splitters with $n$ slots, corresponding to $l_1 = l_2 = \cdots = l_q$. The more general case $l_1 \leq l_2 \leq \cdots \leq l_q$ can be easily realized by allowing beams not to pass *all k n*-slit arrays.

By recalling equation (161), it is easy to compute the probability that a particular program $p_j$ of length $l_j \leq k$ is executed. It is

$$|\langle s_0 s_{1i_1} s_{2i_2} s_{3i_3} \cdots s_{l_j\, i_{l_j}} \mid U \mid s_0 \rangle|^2 = n^{-l_j} \quad . \tag{163}$$

Therefore, there is an inevitable exponential decrease $n^{-l_j}$ in the execution probability.

One possible way to circumvent attenuation would be to amplify the output signals from the beam splitter array. Classically, amplification and copying of bits is no big deal. In quantum mechanics, however, the no-cloning theorem [60, 54, 35, 14] does not allow copying of quantum bits. Any attempt to copy qbits would result in the addition of noise (e.g., from spontaneous emmission processes) and, therefore, in errornous computations.

In summary, the price for speedups of computations originating in quantum parallelism is a corresponding attenuation of the computation probability. In order to compensate for an exponential decrease of execution probability, one would have to *exponentially increase* the number of (bosonic) quanta in the beam paths. This, however, is equivalent to the trivial solution of an arbitrarily complex problem by the introduction of an arbitrary number of classical parallel computers.

We might have to transcend quantum mechanics in order to do better. However, to cite Einstein again, nobody has any idea of how one can find the basis of such a theory.

*A final kaon: Is the question, "where is a rainbow?" a category mistake?*

# Appendices

# A Hilbert space

**D .1 (Field)** *A set of scalars $K$ or $(K, +, \cdot)$ is a* field *if*

***(I)*** *to every pair $a, b$ of scalars there corresponds a scalar $a + b$ in such a way that*

- ***(i)*** $a + b = b + a$ *(*commutativity*);*
- ***(ii)*** $a + (b + c) = (a + b) + c$ *(*associativity*);*
- ***(iii)*** *there exists a unique* zero *element* $0$ *such that* $a + 0 = a$ *for all* $a \in K$*;*
- ***(iv)*** *To every scalar $a$ there corresponds a unique scalar $-a$ such that* $a + (-a) = 0$.

***(II)*** *to every pair $a, b$ of scalars there corresponds a scalar $ab$, called* product *in such a way that*

- ***(i)*** $ab = ba$ *(*commutativity*);*
- ***(ii)*** $a(bc) = (ab)c$ *(*associativity*);*
- ***(iii)*** *there exists a unique non-zero element* $1$*, called* one*, such that* $a1 = a$ *for all* $a \in K$*;*
- ***(iv)*** *To every non-zero scalar $a$ there corresponds a unique scalar $a^{-1}$ such that* $aa^{-1} = 1$.

***(III)*** $a(b + c) = ab + ac$ *(*distributivity*).*

*Examples:* The sets $\mathbb{Q}, \mathbb{R}, \mathbb{C}$ of rational, real and complex numbers with the ordinary sum and scalar product operators "$+, \cdot$" are fields.

**D .2 (Linear space)** *Let $M$ be a set of objects such as vectors, functions, series et cetera. A set $M$ is a* linear space *if*

***(I)*** *there exists the operation of (generalised) "addition," denoted by "+" obeying*

- ***(i)*** $f + g \in M$ *for all* $f, g \in M$ *(*closedness *under "addition");*
- ***(ii)*** $f + g = g + f$ *(*commutativity*);*
- ***(iii)*** $f+(g+h)=(f+g)+h=f+g+h$ *(*associativity*);*
- ***(iv)*** *there exists a* neutral element $\mathbf{0}$ *for which* $f + \mathbf{0} = f$ *for all* $f \in M$*;*
- ***(v)*** *for all $f \in M$ there exists an* inverse $-f$*, defined by* $f + (-f) = \mathbf{0}$*;*

***(II)*** *there exists the operation of (generalised) "scalar multiplication" with elements of the field $(K, +, \cdot)$ obeying*

- ***(vi)*** $a \in K$ *and* $f \in M$ *then* $af \in M$ *(*closedness *under "scalar multiplication");*
- ***(vii)*** $a(f + g) = af + ag$ *and* $(a + b)f = af + bf$ *(*distributive laws*);*

*(viii)* $a(bf) = (ab)f = abf$ *(*associativity*);*

*(ix) There exists a "scalar unit element"* $1 \in K$ *for which* $1f = f$ *for all* $f \in M$.

*Examples:*
*(i)* vector spaces $M = \mathbb{R}^n$ with $K = \mathbb{R}$ or $\mathbb{C}$;
*(ii)* $M = \ell_2, K = \mathbb{C}$, the space of all infinite sequences

$$\ell_2 = \{f \mid f = (x_1, x_2, \ldots, x_i, \ldots),\ x_i \in \mathbb{C},\ \sum_{i=1}^{\infty} |x_i|^2 < \infty\} \quad ,$$

*(iii)* the space of continuous functions, complex-valued (real-valued) functions $M = C(a, b)$ over an open or closed interval $(a, b)$ or $[a, b]$ with $K = \mathbb{C}$ ($K = \mathbb{R}$);

**D .3 (Metric, norm, inner product)**

*A* metric*, denoted by* dist*, is a binary function which associates a distance of two elements of a linear vector space and which satisfies the following properties:*

***(i)*** $\mathrm{dist}(f, g) \in \mathbb{R}$ *for all* $f, g \in M$*;*

***(ii)*** $\mathrm{dist}(f, g) = 0 \Longleftrightarrow f = g$*;*

***(iii)*** $\mathrm{dist}(f, g) \leq \mathrm{dist}(f, h) + \mathrm{dist}(g, h)$ *for all* $h \in M$ *and every pair* $f, g \in M$.

*A* norm $\|\cdot\|$ *on a linear space M is a unary function which associates a real number to every element of M and which satisfies the following properties:*

***(i)*** $\|f\| \geq 0$ *for all* $f \in M$*;*

***(ii)*** $\|f\| = 0 \Longleftrightarrow f = \mathbf{0}$*;*

***(iii)*** $\|f + g\| \leq \|f\| + \|g\|$*;*

***(iv)*** $\|af\| = |a|\,\|f\|$ *for all* $a \in K$ *and* $f \in M$ *(*homogeneity*).*

*An* inner product $\langle \cdot \mid \cdot \rangle$ *is a binary function which associates a complex number with every pair of elements of a linear space M and satisfies the following properties (*$^*$ *denotes complex conjugation):*

***(i)*** $\langle f \mid g \rangle = \langle g \mid f \rangle^*$ *for all* $f, g \in M$*;*

***(ii)*** $\langle f \mid ag \rangle = a\langle f \mid g \rangle$ *for all* $f, g \in M$ *and* $a \in K$*;*

***(iii)*** $\langle f \mid g_1 + g_2 \rangle = \langle f \mid g_1 \rangle + \langle f \mid g_2 \rangle$ *for all* $f, g_1, g_2 \in M$*;*

***(iv)*** $\langle f \mid f \rangle \geq 0$ *for all* $f \in M$*;*

***(v)*** $\langle f \mid f \rangle = 0 \Longleftrightarrow f = \mathbf{0}$.

*Remarks:*

*(i)* With the identifications

$$\|f\| = \langle f|f\rangle^{1/2} \quad , \tag{164}$$

$$\text{dist}(f,g) = \|f-g\| \quad , \tag{165}$$

features & structures are inherited in the following way:

$$M \text{ has an inner product } \underset{\not\Leftarrow}{\Rightarrow} M \text{ has a norm } \underset{\not\Leftarrow}{\Rightarrow} M \text{ has a metric.}$$

*(ii)* The *Schwartz inequality*

$$|\langle f \mid g\rangle| \le \|f\|\|g\| \tag{166}$$

is satisfied.

**D .4 (Separability, completeness)**

*A linear space $M$ is* separable *if, for any $f \in M$ and any $\varepsilon > 0$, there exists at least one element $f_i$ of a sequence $\{f_n \mid n \in \mathbb{N},\ f_n \in M\}$ such that*

$$\|f - f_i\| < \varepsilon \quad .$$

*A linear space $M$ is* complete *if any sequence $\{f_n \mid n \in \mathbb{N},\ f_n \in M\}$ with the property*

$$\lim_{i,j\to\infty} \|f_i - f_j\| = 0$$

*defines a unique limit $f \in M$ such that*

$$\lim_{i\to\infty} \|f - f_i\| = 0 \quad .$$

**D .5 (Hilbert space, Banach space)**

*A* Hilbert space $\mathfrak{H}$ *is a* linear space, *equipped with an* inner product, *which is* separable & complete.

*A* Banach space *is a* linear space, *equipped with a* norm, *which is* separable & complete.

*Example:*

$\ell_2, \mathbb{C}$ [see linear space example *(ii)*] with $\langle f \mid g\rangle = \sum_i x_i^* y_i$.

**D .6 (Subspace, orthogonal subspace)**

*A* subspace $\mathfrak{S} \subset \mathfrak{H}$ *of a Hilbert space is a subset of $\mathfrak{H}$ which is closed under scalar multiplication and addition, i.e., $f,g \in H$, $a \in K \Rightarrow af \in \mathfrak{S}$, $f+g \in \mathfrak{S}$, and which is separable and complete.*

*An* orthogonal subspace $\mathfrak{S}^\perp$ *of $\mathfrak{S}$ is the set of all elements in the Hilbert space $\mathfrak{H}$ which are orthogonal to elements of $\mathfrak{S}$, i.e.,*

$$\mathfrak{S}^\perp = \{f \mid f \in \mathfrak{H},\ \langle f \mid g\rangle = 0, \text{for all } g \in \mathfrak{S}\}.$$

*Remarks:*
*(i)* $(\mathfrak{S}^{\perp})^{\perp} = \mathfrak{S}^{\perp\perp} = \mathfrak{S}$;
*(ii)* every orthogonal subspace is a subspace;
*(iii)* A Hilbert space can be represented as a direct sum of orthogonal subspaces.

**D .7 (Linear function)** *A map* $F : \mathfrak{H} \to K$ *is a* linear function *on* $\mathfrak{H}$ *if*

***(i)*** $F(f+g) = F(f) + F(g)$,

***(ii)*** $F(af) = aF(f)$ *with* $f, g \in \mathfrak{H}$, $a \in K$.

*A linear function is* bounded *if* $|F(f)| \leq a\|f\|$ *with* $a \in \mathbb{R}_+$ *for all* $f \in \mathfrak{H}$.

**T .8 (Dual Hilbert space)** *There exists a one-to-one map between the elements f of a Hilbert space* $\mathfrak{H}$ *and the elements of F of the set* $\mathfrak{H}^\dagger$ *of bounded linear functions on* $\mathfrak{H}$, *such that*

$$F_f(g) = \langle f \mid g \rangle \qquad .$$

*With the operations* ($f, g \in \mathfrak{H}$, $a \in K$)

***(i)*** $F_f + F_g = F_{f+g}$,

***(ii)*** $aF_f = F_{a^* f}$,

***(iii)*** $\langle F_f \mid F_g \rangle^\dagger = \langle g \mid f \rangle$

$\mathfrak{H}^\dagger$ *is the* dual Hilbert space *of* $\mathfrak{H}$.

*Remarks:*
*(i)* $\mathfrak{H} = \{f, g, h, \ldots\}$ and $\mathfrak{H}^\dagger = \{F_f, F_g, F_h, \ldots\}$ are isomorphic; instead of $h \equiv F_h$, one could write $h_F \equiv F$;
*(ii)* $(\mathfrak{H}^\dagger)^\dagger = \mathfrak{H}$.

**T .9 (Isomorphism of Hilbert spces)** *All separable Hilbert spaces of equal dimension with the same field k are isomorphic.*

# B Fundamental constants of physics and their relations

## B.1 Fundamental constants of physics

$c = 2.998 \times 10^8$ m sec$^{-1}$ (velocity of light in vacuum)
$h = 4.136 \times 10^{-15}$ eV sec (Planck's constant)
$\hbar = h/2\pi = 6.582 \times 10^{-16}$ eV sec (Planck's constant/$2\pi$)
$\hbar c = 1.973 \times 10^{-7}$ eV m
$k_B = 8.617 \times 10^{-5}$ eV K$^{-1}$ (Boltzmann's constant)
$e = 1.602 \times 10^{-19}$ coulomb (elementary electron charge)
$\alpha = e^2/\hbar c = 1/137.036$ (fine structure constant)
$\lambda_e = \hbar/m_e c = 3.866 \times 10^{-13}$ m
$\alpha_{\infty,Bohr} = \hbar/m_e e^2 = 0.529$ Å$= 5 \times 10^{-11}$ m (Bohr radius)
$\mu_{Bohr} = e\hbar/2m_e c = 5.788 \times 10^{-9}$ eV gauss$^{-1}$

## B.2 Conversion tables

1 m $^{-1}$ = 1.240 meV ($\times hc$)

0 K = –273° C

1 K = $8.617 \times 10^{-2}$ meV ($\times k_B$)= $6.949 \times 10^{-2}$ m$^{-1}$ ($\times k_B hc$)

1 Å= $10^{-10}$ m

1 eV = $1.602 \times 10^{-19}$ J

## B.3 Electromagnetic radiation and other wave phenomena

| type | frequency [sec$^{-1}$] | energy ($\times h$) [eV] | wavelength (c/frequency) [m] |
|---|---|---|---|
| electric disturbancies (field) | $10^2$ | $4 \times 10^{-15}$ | $3 \times 10^6$ |
| radio | $5 \times 10^5 - 10^6$ | $2 - 4 \times 10^{-9}$ | $6 - 3 \times 10^2$ |
| FM–TV | $10^8$ | $4 \times 10^{-7}$ | 3 |
| radar | $10^{10}$ | $4 \times 10^{-5}$ | $3 \times 10^{-2}$ |
| light | $5 \times 10^{14} - 10^{15}$ | $2 - 4$ | $6 - 3 \times 10^{-7}$ |
| X-ray | $10^{18}$ | $4 \times 10^3$ | $3 \times 10^{-10}$ |
| $\gamma$–radiation | $10^{21}$ | $4 \times 10^6$ | $3 \times 10^{-13}$ |
| cosmic radiation | $10^{27}$ | $4 \times 10^{12}$ | $3 \times 10^{-19}$ |
| elastic lattice vibrations (phonons) | $5 \times 10^{12}$ | $3 \times 10^{-3}$ (10-100 K) | $10^{-9}$ |
| Fermi energy | | $1 - 10$ ($10^4 - 10^5$ K) | |
| plasma frequency | | $5 - 15$ ($10^4 - 10^5$ K) | |

# C Mathematica code for quantum interference

## C.1 Mach-Zehnder interferometer

```
x=a;
x = x/. a -> (b + I c)/Sqrt[2];
x = x/. b -> b Exp[I p];
x = x/. b -> (e + I d)/Sqrt[2];
x = x/. c -> (d + I e)/Sqrt[2];
Print[Expand[x]];
```

## C.2 Mandel interferometer

```
x=a;
x = x/. a -> (b + I c)/Sqrt[2];
```

```
x = x/. b -> eta *d*e;
x = x/. c -> eta *h*k;
x = x/. h -> h Exp[I p];
x = x/. e -> (T* g + I*R*f);
x = x/. h -> (l + I m)/Sqrt[2];
x = x/. d -> (m + I l)/Sqrt[2];
x = x/. g -> k;
(*
x = x/. T -> 1;
x = x/. R -> 0;
x = x/. m -> 0;
x = x/. f -> 0;
*)
Print[Expand[x]];
```

## C.3 Elementary quantum interference device

### Beam splitter

```
 x=i;
 x = x/. a -> a Exp[I alpha + I*beta ];
 x = x/. i -> i Exp[I beta];
 x = x/. a -> (T*e + I*R* d);
 x = x/. i -> (T*d + I*R* e);
 x = x/. d -> d Exp[I varphi];
 x = x/. T -> Cos[omega];
 w = x/. R -> Sin[omega];
 (* Print[Expand[x]]; *)
a22= w/. d -> 0;
a22= a22/. e -> 1;
a21= w/. e -> 0;
a21= a21/. d -> 1;

 x=a;
 x = x/. a -> a Exp[I alpha + I*beta ];
 x = x/. i -> i Exp[I beta];
 x = x/. a -> (T*e + I*R* d);
 x = x/. i -> (T*d + I*R* e);
 x = x/. d -> d Exp[I varphi];
 x = x/. T -> Cos[omega];
 w = x/. R -> Sin[omega];
(* Print[Expand[x]]; *)
a12= w/. d -> 0;
a12= a12/. e -> 1;
```

```
a11= w/. e -> 0;
a11= a11/. d -> 1;

Print["inverse of transition matrix"];
t12=Simplify[{{a11,a12},{a21,a22}}];
Print[t12];

t12i=ComplexExpand[Transpose[Conjugate[ComplexExpand[t12]]]]
```

**Mach-Zehnder**

```
 x=i;
 x = x/. a -> a Exp[I alpha + I*beta ];
 x = x/. i -> i Exp[I beta];
 x = x/. a -> (b + I* c)/Sqrt[2];
 x = x/. i -> (c + I* b)/Sqrt[2];
 x = x/. c -> c Exp[I omega];
 x = x/. b -> (d + I* e)/Sqrt[2];
 x = x/. c -> (e + I* d)/Sqrt[2];
 w = x/. d -> d Exp[I varphi];
 (* Print[Expand[x]]; *)
a22= w/. d -> 0;
a22= a22/. e -> 1;
a21= w/. e -> 0;
a21= a21/. d -> 1;

 x=a;
 x = x/. a -> a Exp[I alpha + I*beta ];
 x = x/. i -> i Exp[I beta];
 x = x/. a -> (b + I* c)/Sqrt[2];
 x = x/. i -> (c + I* b)/Sqrt[2];
 x = x/. c -> c Exp[I omega];
 x = x/. b -> (d + I* e)/Sqrt[2];
 x = x/. c -> (e + I* d)/Sqrt[2];
 w = x/. d -> d Exp[I varphi];
(* Print[Expand[x]]; *)
a12= w/. d -> 0;
a12= a12/. e -> 1;
a11= w/. e -> 0;
a11= a11/. d -> 1;

Print["inverse of transition matrix"];
t12=Simplify[{{a11,a12},{a21,a22}}];
Print[t12];
```

```
t12i=ComplexExpand[Transpose[Conjugate[ComplexExpand[t12]]]];

(* yields Sqrt[not]-gate *)
 y = t12i/. omega -> -Pi/2;
 y = y/. alpha -> -Pi;
 y = y/. beta -> 3Pi/4;
 y = y/. varphi -> 0;
Print[Simplify[y]];
```

# D Recommended reading

## History of quantum mechanics

Jammer [43], Wheeler & Zurek [84]

## Hilbert space quantum mechanics

Feynman, Leighton & M. Sands [32], Harris [40], Lipkin [51], Ballentine [2], Messiah [58], Peres [63], von Neumann [83], Bell [3],

## From single to multiple quanta — “second” field quantization

J. M. Jauch [44], Bogoliubov & Shirkov [11], D. Luriè [53], Itzykson & Zuber [42],

## Quantum interference

Reck, Zeilinger, Bernstein & Bertani [67], Greenberger, Horne & Zeilinger [38], Yurke, McCall & Clauder [86], Campos, Saleh & M. C. Teich [13],

## Copying and cloning of qbits

Milonni & Hardies [60], L. Mandel [54], Glauber [35], Caves [14],

## Context dependence of qbits

Mermin [57],

## Classical versus quantum tautologies

Specker [79],

## Quantum computation

Feynman [33],

### Universal quantum computers

Deutsch [24],

### Universal quantum networks

Feynman [34], Deutsch [25],

### Quantum recursion theory

K. Svozil [81],

### Factoring

Bennett [6], Bernstein & Vazirani [8], Berthiaume & Brassard [9] Černý [16], Shor [76].